\documentclass[10pt,twocolumn,twoside]{IEEEtran}

\usepackage{graphicx} 
\usepackage{float} 
\ifCLASSINFOpdf
\else
\fi

\usepackage{mathtools, cuted}
\usepackage{adjustbox}
\usepackage{amsmath}
\usepackage{amssymb}
\usepackage{hyperref}
\usepackage{graphicx}

\usepackage{subcaption}

\usepackage{nccmath}

\newcommand{\Dtilde}{\tilde{D}}

\newcommand{\defint}[4]{\int_{#1}^{#2}{#3} d{#4}}
\newcommand{\defintrev}[4]{\int_{#1}^{#2}d{#4}{#3}}
\newcommand{\w}{\omega}

\newcommand{\B}{\beta}

\newcommand{\expn}[1]{e^{#1}}

\newcommand{\Ap}{A_p}
\newcommand{\bp}{b_p}
\newcommand{\Bu}{B_u}

\newcommand{\vstapes}{v_{stapes}}
\newcommand{\pconj}{\overline{p}}

\renewcommand{\xi}{x_i}

\newcommand{\ttilde}{\tilde{t}}

\usepackage[usenames, dvipsnames]{color}

\usepackage{ulem}

\newcommand{\trackchanges}[1]{\textcolor{Black}{#1}}
\newcommand{\trackdeletion}[1]{}
\newcommand{\trackaddition}[1]{\textcolor{Black}{#1}}

\usepackage{mathtools}

\usepackage{longtable} 

\usepackage{mathtools}

\newcommand{\appropto}{\mathrel{\vcenter{
  \offinterlineskip\halign{\hfil$##$\cr
    \propto\cr\noalign{\kern2pt}\sim\cr\noalign{\kern-2pt}}}}}

\newcommand{\Pbold}{\mathbf{P}}

\newcommand{\Bcenter}{\B_{peak}}

\newcommand{\CFx}{\text{CF}(x)}

\usepackage{diagbox}

\newcommand{\hGTF}{h_{GTF}}

\usepackage[utf8]{inputenc}

\newcommand{\mytitle}{Rational-Exponent Filters  \\ with Applications to Generalized Exponent Filters}

\newcommand{\myname}{Samiya A Alkhairy }

\newcommand{\myemail}{samiya@alum.mit.edu, samiya@mit.edu}

\newcommand{\myabstract}{We present filters with rational exponents in order to provide a continuum of filter behavior not classically achievable. We discuss their stability, the flexibility they afford, and various representations useful for analysis, design and implementations. We do this for a generalization of second-order filters which we refer to as rational-exponent Generalized Exponent Filters (GEFs) that are useful for a diverse array of applications. We present equivalent representations for rational-exponent GEFs in the time and frequency domains: transfer functions, impulse responses, and integral expressions - the last of which allows for efficient real-time processing without preprocessing requirements. Rational-exponent filters enable filter characteristics to be on a continuum rather than limiting them to discrete values thereby resulting in greater flexibility in the behavior of these filters without additional complexity in causality and stability analyses compared with classical filters. In the case of GEFs, this allows for having arbitrary continuous rather than discrete values for filter characteristics such as (1) the ratio of 3dB quality factor to maximum group delay - particularly important for filterbanks which have simultaneous requirements on frequency selectivity and synchronization; and (2) the ratio of 3dB to 15dB quality factors that dictates the shape of the frequency response magnitude.}

\begin{document}

\title{\mytitle}

\author{\myname

\thanks{\myname  is with MIT, Cambridge, MA, 02139, USA. e-mail: \myemail}}

\markboth{IEEE Transactions on Circuits and Systems I: Regular Papers}{IEEE Transactions on Circuits and Systems I: Regular Papers}

\maketitle

\begin{abstract}
\myabstract
\end{abstract}

\begin{IEEEkeywords}
rational-exponent filters, generalized-exponent filters, Riemann-Liouville integral, auditory filters, fractional-order filters, stability, transfer function, impulse response, quality factor
\end{IEEEkeywords}

\IEEEpeerreviewmaketitle

\maketitle



\section{Introduction}
\label{s:intro}

\subsection{Motivation and Goal}
\label{s:motivation}

Linear time invariant filters are classically represented by rational transfer functions. One generalization of these classical filters is fractional-order filters. Recent interest in fractional-order filters and fractional calculus has increased due to their usefulness for a variety of applications including modeling circuits and systems, control, and filtering signals and noise \cite{podlubny1998fractional, muresan2021review, sang2011neural}. 

Much work on these filters has been dedicated to specific types of fractional-order filters, and especially those that are generalizations of first and second order filters - e.g.,

\begin{equation}
H(s) = \frac{1}{s^{2\alpha} + 2ab s^{\alpha} + b^2}   , \quad   0 < \alpha < 1 \;.
\end{equation}

These include efforts to analyze the behavior (e.g. stability analysis \cite{radwan2012stability}, oscillatory behavior \cite{radwan2008fractional}, analytic solutions \cite{lin2021experimental}) and develop realizations (e.g. state space formulations based on approximations, fractional-order circuit elements) \cite{sarafraz2015realizability, tavakoli2010notes}. 


We are interested in a different generalization of classical filters that we refer to as rational-exponent filters, and that we represent as a base rational transfer function raised to a rational exponent $\Bu$,

\begin{equation}
H(s) =   H_{base}^{B_u}(s)     \quad  \Bu , \in \mathbb{Q}^+  \;.
\end{equation}

The exponent $\Bu$ introduces an additional degree of freedom compared to the base filter $H_{base}(s)$. Further allowing $\Bu$ to be rational and not limiting it to integer values allows for greater flexibility and greater control over the behavior of the filter.

From a theoretical lumped circuit standpoint, non-integer-exponent filters may be thought of as fractional circuits or fractional circuit cascades, in contrast to fractional-order filters that may be represented using fractional circuit elements (e.g. fractional capacitors). Additionally, if a stable system is desired, pole-placement for a rational-exponent filter need only consider the poles of the corresponding base filter, whereas this is a more complicated problem for fractional-order filters. Achieving additional filter design criteria for rational-exponent filters may benefit from those derived for integer-exponent filters \cite{paperB1}.

Our goal is to \trackaddition{(a)} present rational-exponent filters, \trackaddition{(b)} demonstrate the greater flexibility in filter behavior afforded by these filters compared to their integer counterparts, and \trackaddition{(c)} derive and study time and frequency domain representations – that may be useful for realizations as well as further analysis of the filters. We do this for a specific example of rational-exponent filters in which the base filter is second order, and which we introduce in the next section. We note previous efforts that discusses another type of rational-exponent filters in which the base filter is first order \cite{helie2014simulation} \footnote{Though the author refers to these filters as fractional-order filters, we refer to them as rational-exponent filters given their greater similarity to the filters we discuss here compared with existing fractional-order filters.}.

With regards to the various representations of the rational-exponent filter we present here, we note that (1) Some representations have restrictions on filter-exponent whereas this constraint may be reduced in others. (2) Certain representations are best suited for extending the GEFs to nonlinear filters. (3) Yet other representations enable parameterizing the filters in terms of desired sets of time or frequency domain filter characteristics - e.g. rise time, settling time, peak frequency, quality factor, and group delay, thereby providing a great deal of control over the filter behavior without the need for using fixed sets of values (as is traditionally the case for this class of filters), or for optimizing over parameter values for filter design \cite{vecchi2022comparative}. (4) Other representations may be particularly appropriate for real-time signal processing. 

\subsection{Generalized Exponent Filters \trackaddition{with Rational Exponents}}

In this paper, we study rational-exponent filters that are a generalization of second-order filters and are of the form,

\begin{equation}
H(s) = \bigg( \frac{1}{s^2 + 2\Ap s + \Ap^2 + \bp^2} \bigg)^{\Bu}, \quad \Ap,\bp  \in \mathbb{R}^+, \Bu \in \mathbb{Q}^+ \;.
\end{equation}

In the case of unitary exponents, these filters are simply second order filters. We have chosen this form of filters as it is useful for a diverse set of signal processing applications. Filters of this form have previously been introduced for the case of integer exponents ($\Bu  \in \mathbb{Z}^+$): the All-Pole Gammatone Filters (APGFs) \cite{lyon1997all}, and the Generalized Exponent Filters (GEFs) \cite{alkhairy2019analytic, paperB1}. As $\Bu$ has no upper limit, these filters allow for an expanded range of signal processing behavior compared with the corresponding base filter which is second order. However, the filter behavior achievable with the integer-exponent filters is discrete and not on continuum - thereby limiting accessible filter behavior.

APGFs and GEFs are similar to one another in form, but use different notation, and derivations. Either notation may be used, but we have chosen to use and extend the GEF notation for three main reasons:
\begin{itemize}
\item The GEFs may be parameterized in terms of characteristics such as peak frequency, quality factors, and ratio of quality factor to normalized group delay rather than the generic filter constants \cite{alkhairy2022cochlear, paperB1}. This allows us to design filters based on these characteristics and to control the specifications of the filter easily without requiring manual tuning or optimization. 
\item \trackaddition{For a subset of parameter values, the aforementioned filters mimic signal processing by the auditory system.} We have previously tested the GEFs and shown that they are able to mimic cochlear signal processing of low-level \trackaddition{inputs} \cite{alkhairy2019analytic}. This is desired for various signal processing applications including those unrelated to audio signals.
\item GEFs may be used for a variety of signal processing applications as well as the scientific study of the cochlea due to its derivation \cite{alkhairy2019analytic}. 
\end{itemize}


In the context of applications, GEFs are mostly used in the form of parallel filterbanks with each constitutive filter in the filterbank having a different peak frequency. Potential applications for GEFs include: multiplexers \cite{galbraith2008cochlea},   designing rainbow sensors, \cite{marrocchio2021waves}, analyzing seismic signals \cite{jiang2020automatic}, underwater sound classification \cite{zeng2014underwater}, cochlear implants \cite{yao2002application}, and hearing aids \cite{zhang2016intelligent}. For most of these applications - including those unrelated to audio signals, it is desirable that the filter constant values be chosen such that the filterbank mimics signal processing of the mammalian cochlea in the auditory system.

\subsection{Objectives}
\label{s:objectives}

Our objective is to present rational-exponent GEFs in order to achieve a continuum of possible filter characteristics and behavior without added complexity in stability and causality analyses incurred with fractional-order generalizations of classical filters. We also derive and analyze various time and frequency domain representations for the rational-exponent GEFs that afford a great deal of flexibility when developing realizations of these filters.

We are interested in rational-$\Bu$ GEFs as they allow for a greater variety of signal processing behavior that can be accomplished compared to the restrictive integer-$\Bu$ filters or the even more restrictive classical rational transfer functions (such as second order filters). While not the topic of this work, an additional benefit of allowing for non-integer-$\Bu$ GEFs is the ability to better study cochlear mechanisms underlying function.

\subsection{Domains and Representations}
\label{s:scope}


The filter expressions are derived in the continuous frequency and continuous time domains. Our expressions are a function of independent variables - time, frequency, or Laplace-s, or their normalized counterparts. We abuse notation and take liberties in referring to independent variables as frequency or Laplace domains. As this paper is not primarily concerned with implementations, we are not concerned with discretizations and transformations to corresponding digital filters here. However, we note that the time domain representations may directly be transformed into their digital counterparts using standard methods towards digital realizations of the filter.

We derive the various time and frequency domain representations as a means to best fulfill our objective of deriving realizable rational-$\Bu$ GEFs. Having a variety of representations also enables implementational flexibility which is a desirable feature of filters. The ability to implement GEFs in the most suitable way for a particular application may involve designing computer software, or analog or digital hardware architectures. In translating the non-integer-exponent representations into implementations, we might additionally seek inspiration from previous work on realizations for fractional-order filters or their approximations \cite{ali2013fractional, sarafraz2015realizability, tavakoli2010notes}. The integer-exponent instances of GEFs may benefit from existing implementations for related filters \cite{kim2018open, summerfield1992asic, van2003digital, takeda2017novel}.

Importantly, we note that for several  applications, simple direct software implementations are most suitable. The representations we present here for rational-exponent GEFs may relatively directly be used in software and embedded software solutions. Additional work is necessary for developing analog and digital hardware implementations based on some of the representations we derive for non-integer-$\Bu$ GEFs.

\subsection{Organization}

To achieve our objective of developing rational-exponent filters - and specifically those that are generalizations of second order filters (GEFs), we present filter representations in the time and frequency domains and analyze them. We first introduce the reader to the rationale behind deriving each of the representations in this paper (in section \ref{s:representationsNormalizedFandT}). We then introduce normalized frequency and normalized time domains in which we derive the representations. We present the transfer function representation in section \ref{s:TF} and discuss the stability and causality of the rational-exponent filters.

We also illustrate the continuum of behavior accessible due to relieving the integer-$\Bu$ constraint. We then use the transfer function representation to derive the impulse response representation (section \ref{s:impulseResponse}) and the ODE representation (section \ref{s:tODE}). Our last representation is in the form of integral expressions (section \ref{s:integralExpress}) derived from the ODE representation using differential operator theory and extrapolated to non-integer $\Bu$. We test the various representations and assess their equivalence using certain inputs in section \ref{s:equivRep}, and finally provide our conclusions and future directions in section \ref{s:conclusionFutureDir}.

\section{Overview of Representations}
\label{s:representationsNormalizedFandT}

In this section, we provide an overview of the equivalent filter representations presented in this paper. In addition to the the benefits particular to each representation individually, we note that collectively having multiple representations enables implementational flexibility. The representations are in continuous time and continuous frequency. The time domain representations can directly be translated into their digital counterparts using standard methods. We note that our derivation of the representations for rational-exponent GEFs can be used to guide the derivations for other rational-exponent filters as well.

\subsubsection{Transfer functions}

Integer-exponent GEFs were previously presented and derived in the frequency domain \cite{alkhairy2019analytic}. Here we extend our study of the transfer functions to rational-exponent GEFs. The transfer function representation is the most useful for studying the causality and stability of rational-$\Bu$ GEFs. Additionally, it is useful for filter design paradigms in which it is desirable to achieve certain criteria on frequency domain filter characteristics, such as values of peak frequency and bandwidth. These can be achieved using the transfer function representation \cite{paperB1}.

\subsubsection{Impulse response}

The impulse response representation is useful for parameterizing the filter based on time-domain (specifically impulse response or step response) characteristics such as tonal frequency, rise time, and settling time. This allows for time-domain characteristics-based filter design. We derive the impulse response for integer and half-integer-$\Bu$ GEFs. The impulse response is useful for convolution-based implementations \cite{baranowski2016digital}. We also provide an approximation of the rational-exponent GEF as an extrapolation of a Gammatone Filter (GTF) which is quite powerful as it enables extending existing architectures towards implementing the filter. Both the impulse response and transfer function representations are appropriate for choosing filter constant values for which the GEFs mimic natural signal processing in mammals (which is quite desirable for most applications). This is due to the fact that most experimental measurements are in forms of transfer functions and impulse responses \cite{charaziak2023estimating, van2003cochlear}.

\subsubsection{ODEs}
We derive an ordinary differential equation representation that can be solved with a given set of initial conditions. The ODE representation allows for direct extensions for nonlinear filters and for introducing delays. The ODE representation (extrapolated to non-integer-$\Bu$ cases in form only) also enables us to derive the integral formulation. While the ODE representation is presented only for integer-$\Bu$ GEFs, it is quite appropriate for simple and direct software implementations (and embedded systems) and is direct in the sense that it does not require taking any transforms or zero padding and can be done in real-time. Consequently, the ODE representation may be the most suitable for several of the feature-extraction and classification applications and perceptual studies for which the primary requirement is having an easy, direct, software implementation. Additionally, state space representations may be used to derive equivalent filters (e.g. orthonormal ladder filters) for which circuit design may be easily automated.

\subsubsection{Integral Expressions}

We derive integral representations for rational-$\Bu$ GEFs. The integral representations are formulated to assume zero initial conditions. The integral representation is parameterized by the filter exponent and hence, importantly, does not require constructing a different filter for each choice of filter exponent. Notably, the integral formulation allows for real-time processing without the need for preprocessing such as zero padding or transforms and may particularly be appropriate for integrator-based realizations.

\trackaddition{In the following sections, }we introduce the independent variables for representations and proceed to derive and analyze the aforementioned formulations, and also discuss causality and stability. 


\section{Normalized Frequency}
\label{s:normFrq}

Prior to introducing the transfer function, we discuss our choice of independent variable here. As GEFs are typically used in the form of filterbanks with each filter peaking at a different frequency, it is most appropriate to present the filter frequency responses in a manner which allows us to denote both frequency and peak frequency.

Consequently, for frequency domain representations (transfer functions), we consider a filter with peak frequency $\w_i$ to have a transfer function $H_i(\w)$. Alternatively, we may assign a filter in a filterbank to a fictitious point in space, $x_i$ - corresponding to a location along a fictitious cochlea, with a specified peak (characteristic) frequency, CF$(x_i)$. For a set of filters, this results in $H(x,\w)$. If we may express $H(x,\w)$, in terms of a single independent variable, $\B$, presented in \cite{alkhairy2019analytic}, as $H(x,\w) = \mathcal{H}(\B(x,\w))$, we consider $H$ to be `scaling symmetric'. 

Consequently, we define our expressions in the \textit{normalized} \trackaddition{frequency or} angular frequency domain, $\B$, rather than in $f$ or $\w$, 

\begin{equation}
    \B \triangleq \frac{f}{\CFx} \;,
\end{equation}

where $\CFx$ is the characteristic frequency (or peak frequency) map \cite{alkhairy2019analytic}. We define the associated $s$ which - for the purely imaginary case, takes on the form,

\begin{equation}
    s(\B) = i\B \;.
\end{equation}


The input to the filters is denoted by $v_s(\w)$ \footnote{the notation we use for the input is that of the stapes velocity due to the mechanistic origins of GEF}, and the outputs are denoted by $P(x,\w)$ \footnote{in the cochlear mechanistic model, this corresponds to differential pressure across the Organ of Corti}.

We define \trackchanges{a scaling symmetric $\mathbf{P}(s)$ which is a transfer function normalized to the gain constant}  as,

\begin{equation}
    \mathbf{P}(s) = \frac{P(x,\w)}{C(x) v_{st}(\w)} \;,
    \label{eq:Ps}
\end{equation}

and provide the expressions for this transfer function in section \ref{s:TF}.


\section{Transfer Functions}
\label{s:TF}

\subsection{Transfer Function Representation}

The transfer functions for integer-exponent GEFs were previously derived in \cite{alkhairy2019analytic} where we have shown a wider array of signal processing behavior compared to the GEF base filter which is a second order filter. In this section, we extend the filter to the rational-exponent case. The transfer functions for GEFs are,
\begin{equation}
\begin{aligned}
    \Pbold(\B(x,\w)) & \triangleq \frac{P(x,\w)}{C v_{st}(\w)} = \Big((s-p)(s-\bar{p})\Big)^{-\Bu} \\
    & = \Big( \frac{1}{s^2 + 2\Ap s + \Ap^2 + \bp^2 } \Big)^{\Bu}
    \label{eq:Pbold}
\end{aligned}
\end{equation}

where $\Pbold$ at a particular location is the transfer function for the GEF with peak frequency $\textrm{CF}(x)$. Motivated by potential applications, we are primarily interested in cases where the choice of filter constant values result in a bandpass filter, and  \trackaddition{especially the subset with constant values} that mimic auditory signal processing \trackchanges{as} is desirable for a diverse set of applications.


In the expression, \trackaddition{the pair of repeated complex conjugate poles is} $p = i\bp - \Ap$ and  $\pconj = -i\bp - \Ap$. The filter constants $\Ap, \Bu, \bp$ take on real positive values. If the filters are sharply tuned, $\Ap^2 + \bp^2 \approx \bp^2$, and if the peak frequency occurs at $\CFx$ (i.e. $\Bcenter = 1$), then  $\bp \approx 1$. In this case, we may simplify the above expression to $\Pbold = \Big( \frac{1}{s^2 + 2\Ap s + 1} \Big)^{\Bu}$ which is parameterized by only two constants - $\Ap$ and $\Bu$. The filters are of exponent $\Bu$.

Clearly, the GEFs are second-order filters raised to a power $\Bu$. For cases in which $\Bu$ is an integer, the equation results in a rational transfer function with $\Bu$ pairs of complex conjugate poles. Under these conditions, the linear filter is of an infinite impulse response (IIR) system that is stable and realizable.

\subsection{Stability and Causality of Rational Exponent-Filters}

In this section, we demonstrate that the causality and stability analyses of rational-exponent filters are no more complicated than those of their base filters - in contrast to fractional-order filters. As $C(\w) = C \in \mathbb{R}$, a real input will result in a real output even if the exponent, $\Bu$, is a non-integer number. We demonstrate BIBO stability of rational-exponent filters by making the following observation:

For the case of $\Bu = \frac{m}{n}$ where $m, n \in \mathbb{Z}^+$ and are finite, sequential processing by $n$ filters is equivalent to a stable $m^{th}$ \trackchanges{exponent} system because the poles are in the left half plane. Consequently, each constitutive filter of \trackchanges{exponent} $\Bu$, where $\Bu$ is a positive rational number, must be stable.

Mathematically, we may describe the argument as follows: A cascade of $n$ GEFs is equivalent to a cascade of $m$ base filters,

\begin{equation}
    \Pbold^n(s) = H_{base}^m(s) \;.
    \label{eq:eq8}
\end{equation}

\trackchanges{The} transfer functions on both sides of the above equations are raised to integer exponents. If the base filter is BIBO stable, then a filter composed of a cascade of these filters ($H_{base}^m$) must result in a BIBO stable filter. Consequently, \trackaddition{based on \ref{eq:eq8},} its equivalent, $\Pbold^n$, describes a BIBO stable filter. Finally, since the filter corresponding to $\Pbold^n$ is composed of $n$ identical filters, then each GEF with TF $\Pbold$ must itself also be BIBO stable.

The causality of rational-exponent filters also arises from a similar argument, and is also apparent from the various time-domain representations derived in the subsequent sections. While we expect these properties to hold for the more general case of $\Bu \in \mathbb{R}^+$ due to continuity, we are not concerned with this extension. This is because, \textit{practically}, there is nothing to be gained in terms of finer tuning  of signal processing behavior by moving from rational to real $\Bu$. Therefore, we do not discuss the \trackaddition{causality and} stability  of real-exponent filters. 

Our above analysis regarding causality and stability of rational-exponent filters are not restricted to GEFs. Unlike fractional-order filters which require a more involved stability analysis \cite{radwan2012stability}, the  only condition for the stability and causality of any rational-exponent filter is that the base filter is a rational transfer function that itself is stable and causal. Accordingly, filter design and pole placement for a stable non-integer exponent filter is particularly simple for cases where the base filter is up to fourth order (in which case there are closed form expressions relating filter coefficients and poles).

The rational-exponent filters do not have constraints on the maximum value of the exponent (and hence filter order) thereby enabling a great \textit{range} of filter behavior. This is in contrast to existing fractional-order filters which are generally limited to have a maximum filter order of two.

\subsection{Flexibility Afforded by Non-Integer Exponents}

Integer-exponent GEFs display a large \textit{range} of behavior as there is no maximum limit on the value of $\Bu$. Here, we show that extension to rational-exponents results in a \textit{continuum} of possible filter behavior.

To do so, we first analyze the dependence of the behavior of the GEF TFs on the three filter constants, $\Ap, \bp, \Bu$. We find that:
\begin{itemize}
    \item The peak normalized frequency is $\bp$ 
    \item Group delay increases proportionally with $\Bu$ and inversely with $\Ap$
    \item Bandwidth increases proportionally with $\Ap$ and decreases with $\Bu$
\end{itemize}

\begin{figure*}[!ht]
    \centering
    \begin{subfigure}[t]{0.5\linewidth}
        \centering
        \includegraphics[width=\textwidth]{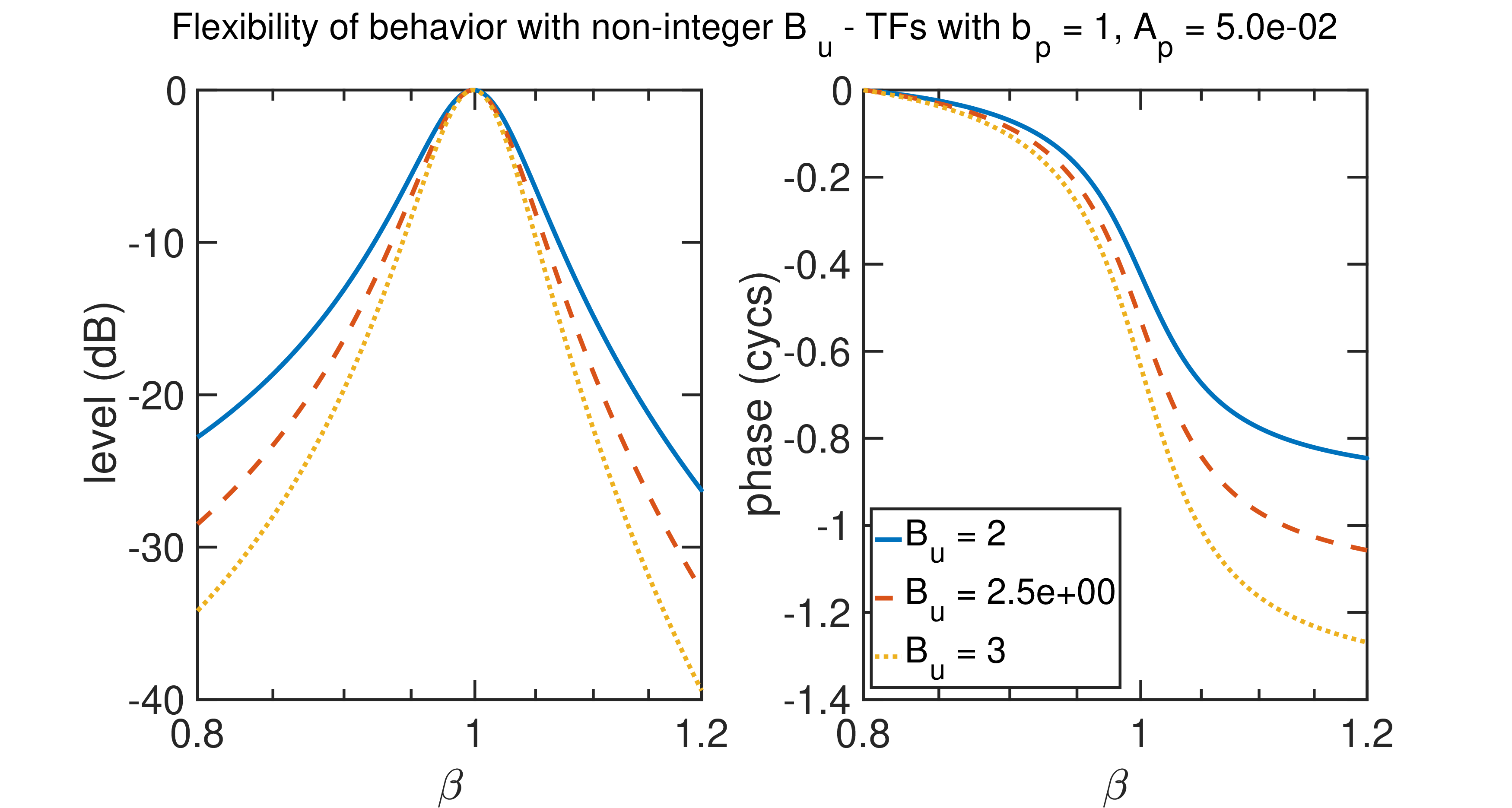}
        \caption{$\Ap = 0.05, \Bu \in \{2,2.5,3\}$}
    \end{subfigure}%
    ~ 
    \begin{subfigure}[t]{0.5\linewidth}
        \centering
        \includegraphics[width=\textwidth]{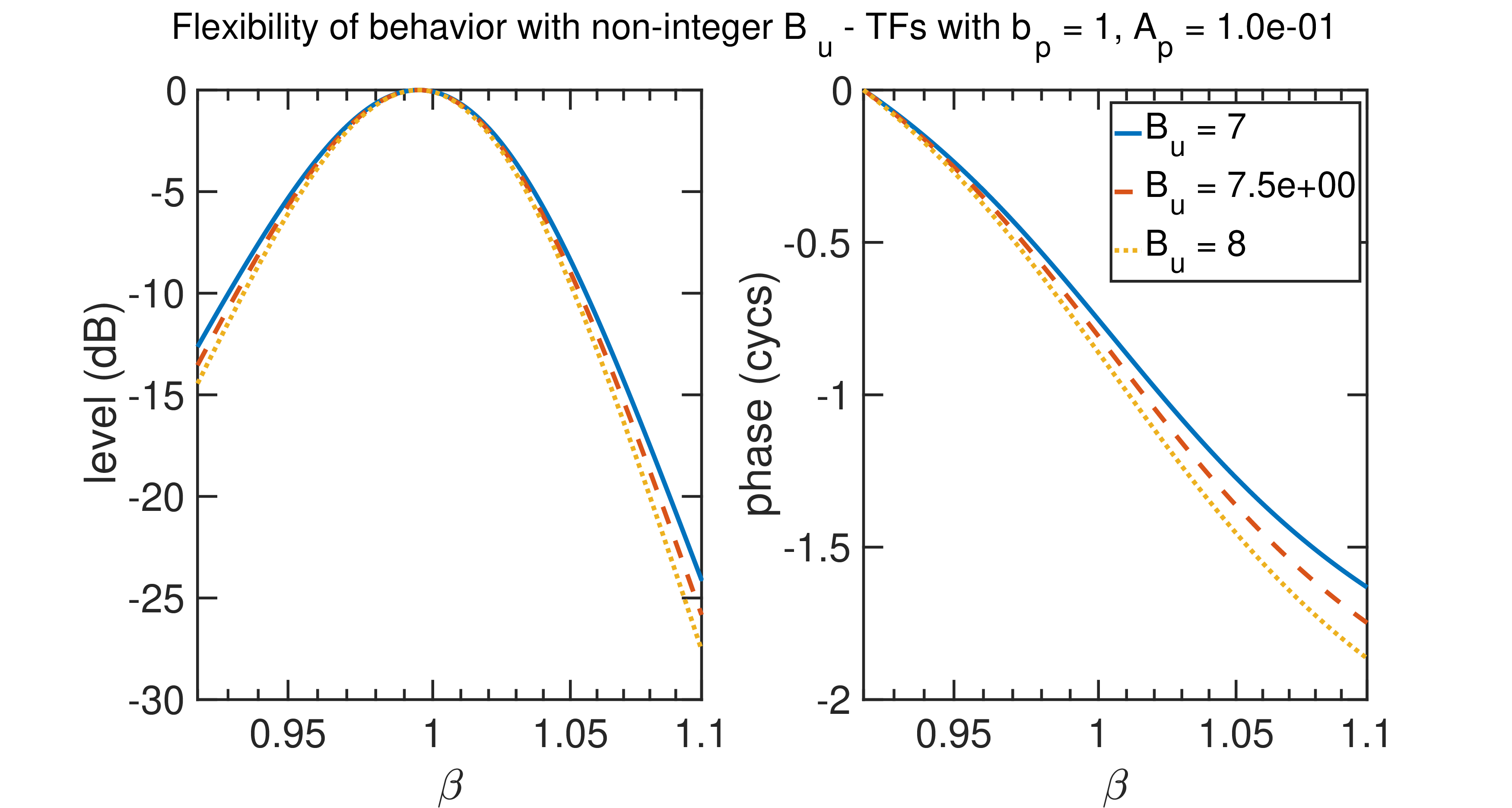}
        \caption{$\Ap = 0.1, \Bu \in \{7,7.5,8\}$}
    \end{subfigure}
    \caption{\textbf{Allowing for non-integer $\Bu$ enables greater variety of behavior:} Figure shows Bode plots of GEFs with integer and half-integer $\Bu$. The two sets of figures are generated with different values for the filter constant $\Ap$ and different ranges for $\Bu$. The left set of figures is more sensitive to changes in $\Bu$ in non-integer increments, illustrating that the flexibility in signal processing behavior afforded by relieving the integer constraint depends on the values of $\Ap$ and $\Bu$. In these plots, the magnitude (in dB) is plotted in reference to the magnitude at the peak, and the phase (in cycles) is in reference to the first phase shown.}
    \label{fig:TF22537758}
\end{figure*}

\begin{figure*}[!ht]
    \centering
    \begin{subfigure}[t]{0.45\linewidth}
        \centering
        \includegraphics[width=\textwidth]{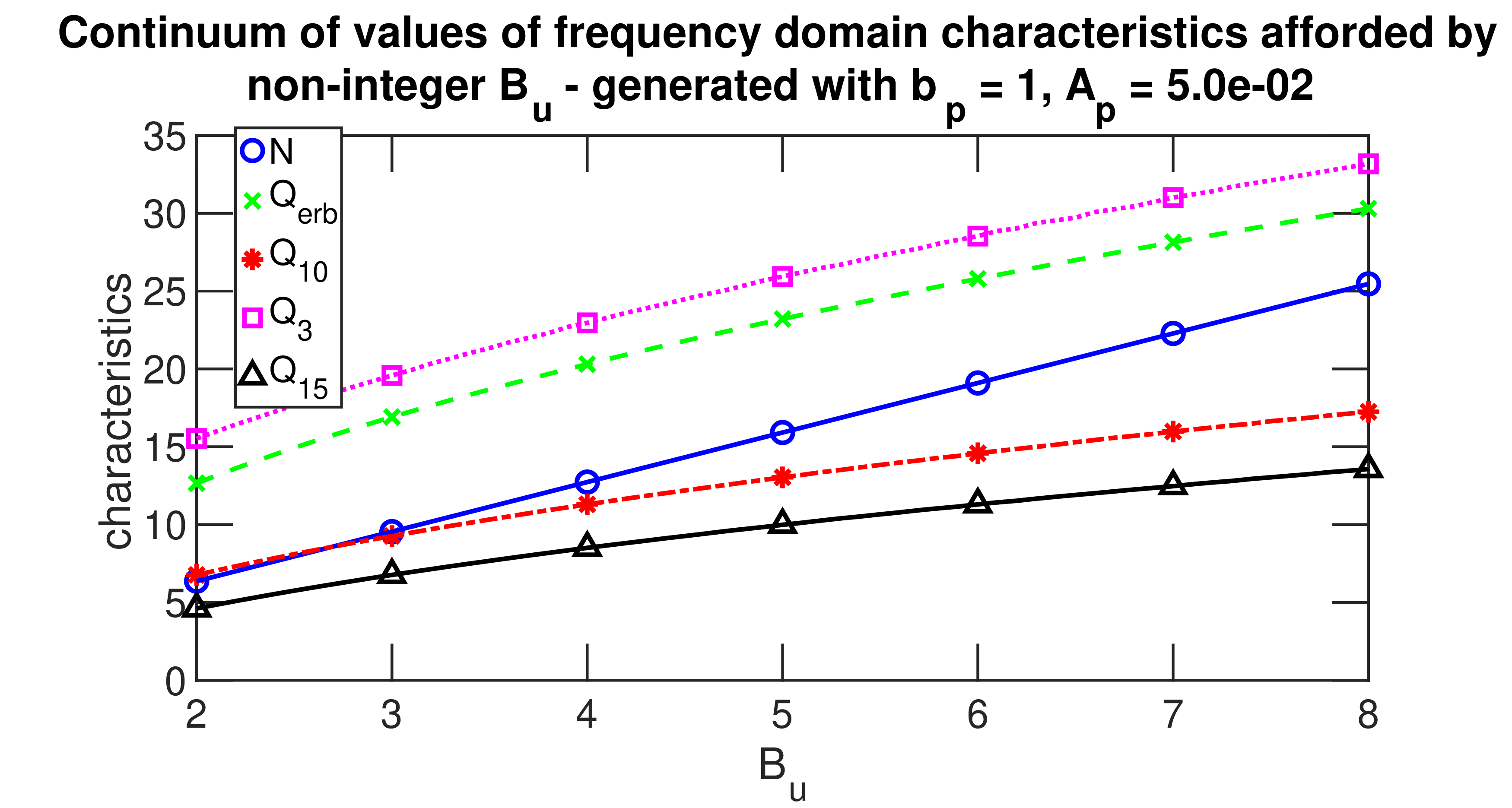}
        \caption{Characteristics are plotted as a function of continuous/rational-$\Bu$ (lines) and as a function of discrete/integer-$\Bu$ (markers). These characteristics are: $Q_{erb}$ which is the quality factor generated from the equivalent rectangular bandwidth $Q_{erb} = \frac{\B_{peak}}{\textrm{ERB}_{\B}}$; $ Q_{15}, Q_{10}, Q_{3}$ which are the quality factors computed using 15, 10, and 3 dB bandwidths respectively; and $N$ which is the maximum normalized group delay, $N = -\frac{1}{2\pi} \max ( \frac{d\textrm{phase}\{\Pbold\}}{d\B}$ )}
    \end{subfigure}%
    ~ 
    \begin{subfigure}[t]{0.45\linewidth}
        \centering
        \includegraphics[width=\textwidth]{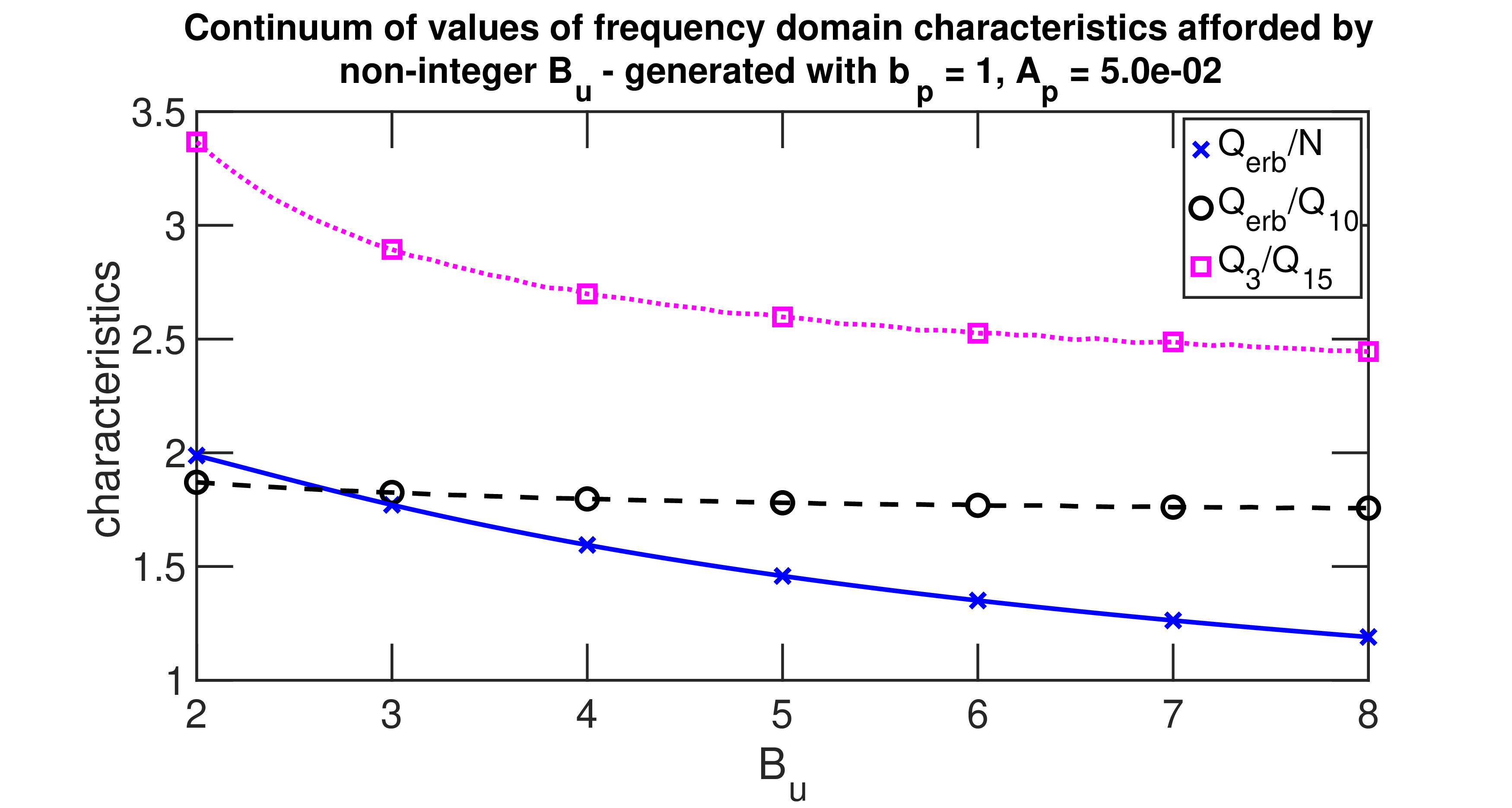}
        \caption{The ratio of quality factor to group delay $\frac{Q_{erb}}{N}$ (which relates frequency selectivity to delay) shows a great degree of \trackchanges{sensitivity to the value of} $\Bu$. The figure also shows the ratio of various quality factors to one another \trackaddition{as a function of $\Bu$}.}
    \end{subfigure}
    \caption[\textbf{Allowing for rational, rather than only integer, $\Bu$ provides greater flexibility in behavior:} The figures show the continuum of behavior (quantified by TF characteristics) that may be accessed once we \trackchanges{allow for rational-$\Bu$}(lines), as opposed to the discrete behavior accessible via the restricted integer-$\Bu$ cases (markers). The left figure shows the continuum of behavior as a function of (practically) continuous $\Bu$ for simple filter characteristics ($N, Q_{erb}, Q_{10}, Q_3, Q_{15}$) while holding $\Ap$ constant. The right figure shows the continuum of behavior seen from the perspective of compound characteristics that practically only depend on $\Bu$ ($\frac{Q_{erb}}{N}, \frac{Q_{erb}}{Q_{10}}, \frac{Q_3}{Q_{15}}$ ).]{\textbf{Allowing for rational, rather than only integer, $\Bu$ provides greater flexibility in behavior:} The figures show the continuum of behavior (quantified by TF characteristics) that may be accessed once we \trackchanges{allow for rational-$\Bu$}(lines), as opposed to the discrete behavior accessible via the restricted integer-$\Bu$ cases (markers). The left figure shows the continuum of behavior as a function of (practically) continuous $\Bu$ for simple filter characteristics ($N, Q_{erb}, Q_{10}, Q_3, Q_{15}$) while holding $\Ap$ constant. The right figure shows the continuum of behavior seen from the perspective of compound characteristics that practically only depend on $\Bu$ ($\frac{Q_{erb}}{N}, \frac{Q_{erb}}{Q_{10}}, \frac{Q_3}{Q_{15}}$ )\footnotemark.}
    \label{fig:charContinuum}
\end{figure*}

Figure \ref{fig:TF22537758} shows the Bode plots for various values of $\Bu$ and illustrates the flexibility in behavior afforded by relieving the integer restriction. Figure \ref{fig:charContinuum} quantifies this by demonstrating the continuum of filter characteristics afforded by having rational $\Bu$. For instance, the plot for the ratio of quality factor to group delay (which is practically purely a function of $\Bu$ \cite{alkhairy2022cochlear})  shows that reducing the integer-$\Bu$ constraint to a rational one, allows for accessing a continuum of this ratio rather than the discrete values we are limited to in the integer-exponent case. Greater flexibility in controlling this ratio is particularly important for filterbanks (the primary configuration in which such filters are used) where frequency selectivity and synchronization requirements are important. Similarly, allowing for non-integer values of $\Bu$ results in a continuum of values for the ratio of the 3dB quality factor to 15dB quality factor which controls the shape of the magnitude of the filter response. These observations demonstrate the fact that non-integer-$\Bu$ GEFs have a greater set of accessible behaviors than those restricted to integer $\Bu$.

\footnotetext{\trackchanges{Strictly speaking $\frac{Q_{erb}}{N}$ also depends on $\bp$ but this is generally fixed to unity so that the peak occurs at the selected peak frequency, $\CFx$.}}

\subsection{Properties}

As described above, the TF representation is the appropriate GEF representation for studying stability and causality. The GEF is BIBO stable and causal if the underlying base filter is BIBO stable and causal. Clearly, this analysis (fundamental to the design of filters) is particularly simple for GEFs and other rational-exponent filters in contrast to fractional-order filters which are difficult to characterize in terms of causality and stability.

As mentioned previously, it is desirable for many applications that the GEFs mimic signal processing of the auditory system. The TF representation is ideal for testing the filter against data and for estimating values of filter constants due to the nature of much of the reported data from cochlear experiments \cite{van2003cochlear}.

Additionally, the simplicity of the form of the TF representation allows us to derive parameterizations of GEFs in terms of desired filter characteristics such as peak frequency, quality factor, and maximum group delay and enables novel methods for characteristic-based filter design \cite{alkhairy2022cochlear}. This is quite desirable for an array of signal processing applications \footnote{It is also useful to auditory scientists seeking to understand the relationship between observed function and underlying mechanisms in the cochlea.}.

\begin{figure}[htbp]
    \centering
    \includegraphics[width = \linewidth]{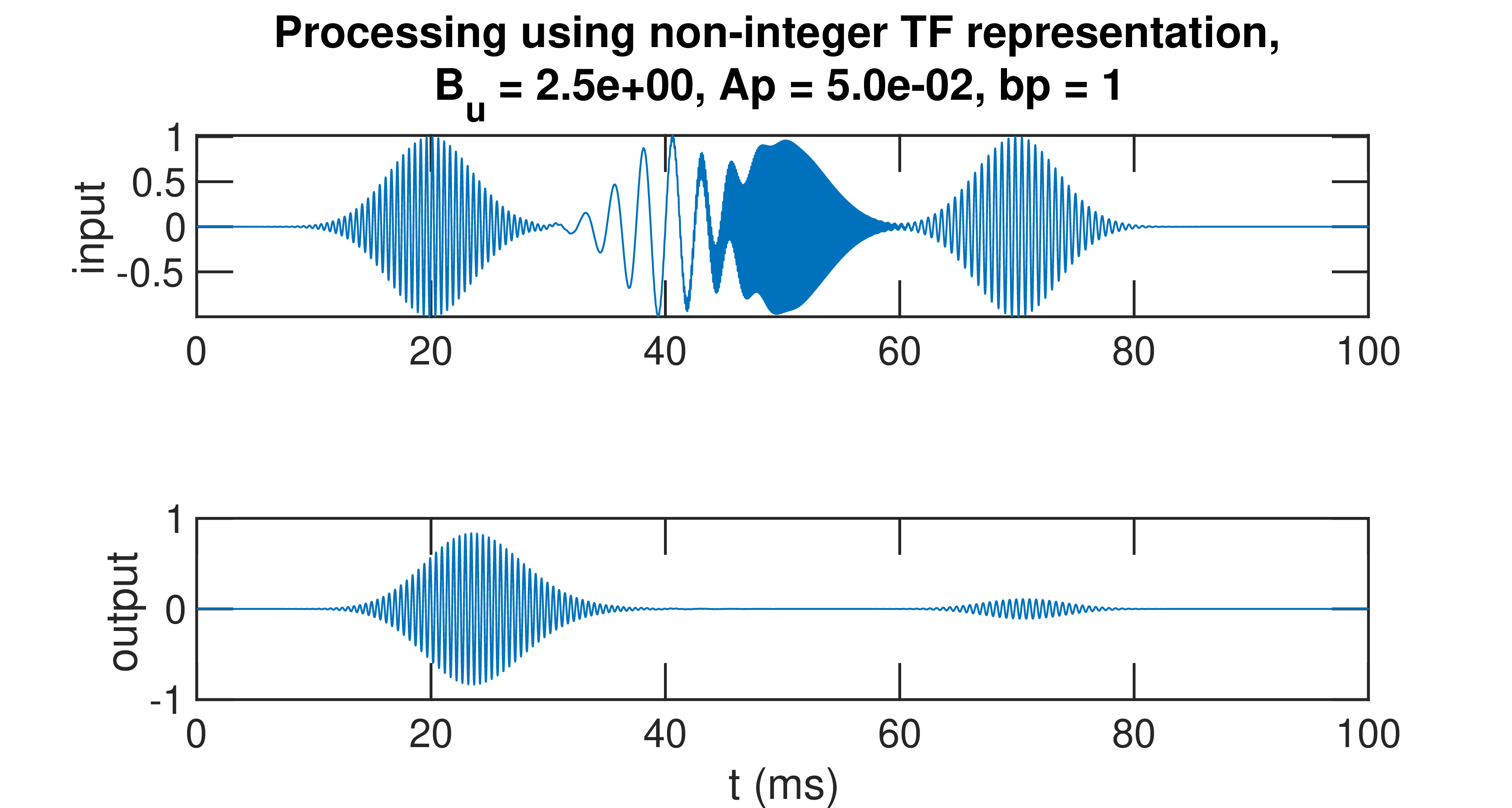}
    \caption{\textbf{Software implementations of the non-integer-$\Bu$ TF representation can be used to process signals:} The output (bottom panel) of a filter with peak frequency $\CFx =$ 2 kHz is computed using a software implementation of the TF representation using a non-integer-$\Bu$ value of $2.5$. The output is generated in response to  
    a input signal (top panel) which is composed of tone pips and is of the form $\vstapes = \sum\limits_{i=0}^3 e^{-(t-t_i)^2/T^2} sin(2\pi f_i t)$, where $T=5, f_0 = \textrm{CF}, t_0 = 20, f_1 = 5 \textrm{CF}, t_1 = 50, f_2 = \frac{7}{8}\textrm{CF}, t_2 = 70, f_3 = \frac{1}{5}\textrm{CF}, t_3 = 40$ with frequency in kHz, and time in ms. As expected, the filter responds maximally to frequency components closest to its $\CFx$. This figure illustrates realizability of software implementations using the TF representation with non-integer-$\Bu$.}
\label{fig:TFimplementability}
\end{figure}

Architectures for analog and digital hardware may easily be developed based on the TF representation for the case of integer $\Bu$ - e.g as delay equations in the form, $x[n] = \sum\limits_{k=0}^{M} a[k] x[n-k] - b y[n]$ due to the all-pole nature of $\Pbold$. We may especially leverage existing work on implementations of the APGF. In figure \ref{fig:TFimplementability}, we demonstrate the ability to computationally process signals using TFs with non-integer-$\Bu$ and hence the ability to use it for software realizations. Future work may involve developing analog and digital realizations based on the non-integer-$\Bu$ TFs (or other representations) which may benefit from approximation and truncation methods and existing work on implementations of fractional-order filters of the classical form. 

\section{Normalized Time}
\label{s:normTime}

For $C$ of equation \ref{eq:Ps} independent of $\w$, our filter has explicit time domain representations that are derived and studied in the following sections. Here, we introduce the independent variable - normalized time, and other time domain variables.

We use $u(t)$ to denote the time domain counterpart of the input $v_{st}(\w)$,
\begin{equation}
        v_{st}(\w) \xleftrightarrow{} u(t) \;.
\end{equation}

\trackaddition{To avoid confusion with the pole $p$ of the TF, } we use $q(x,t)$ to denote the time domain output (relative to $C$) in response to an input,

\begin{equation}
        \frac{P(x,\w)}{C(x)} \xleftrightarrow{} q(x,t) \;,
\end{equation}

As frequency is normalized, we define a scaled version of time, 

\begin{equation}
    \ttilde(t,x) \triangleq 2\pi\text{CF}(x) t \;,
    \label{eq:introduceTtilde}
\end{equation}

to formulate the time domain representations. For instance, we denote the impulse response of the normalized output (normalized to $C$) as follows,

\begin{equation}
\begin{aligned}
    \frac{P(x,\w)}{C(x) v_{st}(\w)} & = \Pbold(s) = \Pbold \bigg( \frac{\hat{s}}{2 \pi \CFx} \bigg)  \\
     \mapsto g(x,t) & = 2\pi \CFx h(\ttilde) = 2\pi \CFx h(2\pi\CFx t) \;,
     \label{eq:P2g}
\end{aligned}
\end{equation}

where we have made use of the scaling property of Laplace transforms. 

The time domain representations derived in this paper - (a) impulse response, (b) ODE, and (c) integral representations, are in continuous time.


\section{Impulse Responses}
\label{s:impulseResponse}



Here we derive and study the first of the time-domain representations - the impulse response. The $h(\ttilde)$ of the normalized impulse response, $g = 2\pi\CFx h(\ttilde)$ (of equation \ref{eq:P2g}) can be expressed as a function of scaled time, $\ttilde$, introduced in equation \ref{eq:introduceTtilde}. We also define, purely for the purposes of compactness,

\begin{equation}
    \ttilde_b \triangleq \bp \ttilde \;.
\end{equation}

We begin by deriving impulse responses for integer-$\Bu$ GEFs, then do so for half-integer-$\Bu$ cases. For integer $\Bu$ we have a restriction of $\Bu > 1$ in order to achieve impulse responses for which the envelope can grow then decay to mimic cochlear signal processing as is desirable for most applications. For half-integer $\Bu$ this restriction is further refined to $\Bu \geq \frac{3}{2}$.

\subsection{Integer $\Bu$: Impulse Response Limitations}

\begin{figure}[htbp]
    \centering
    \includegraphics[width = \linewidth]{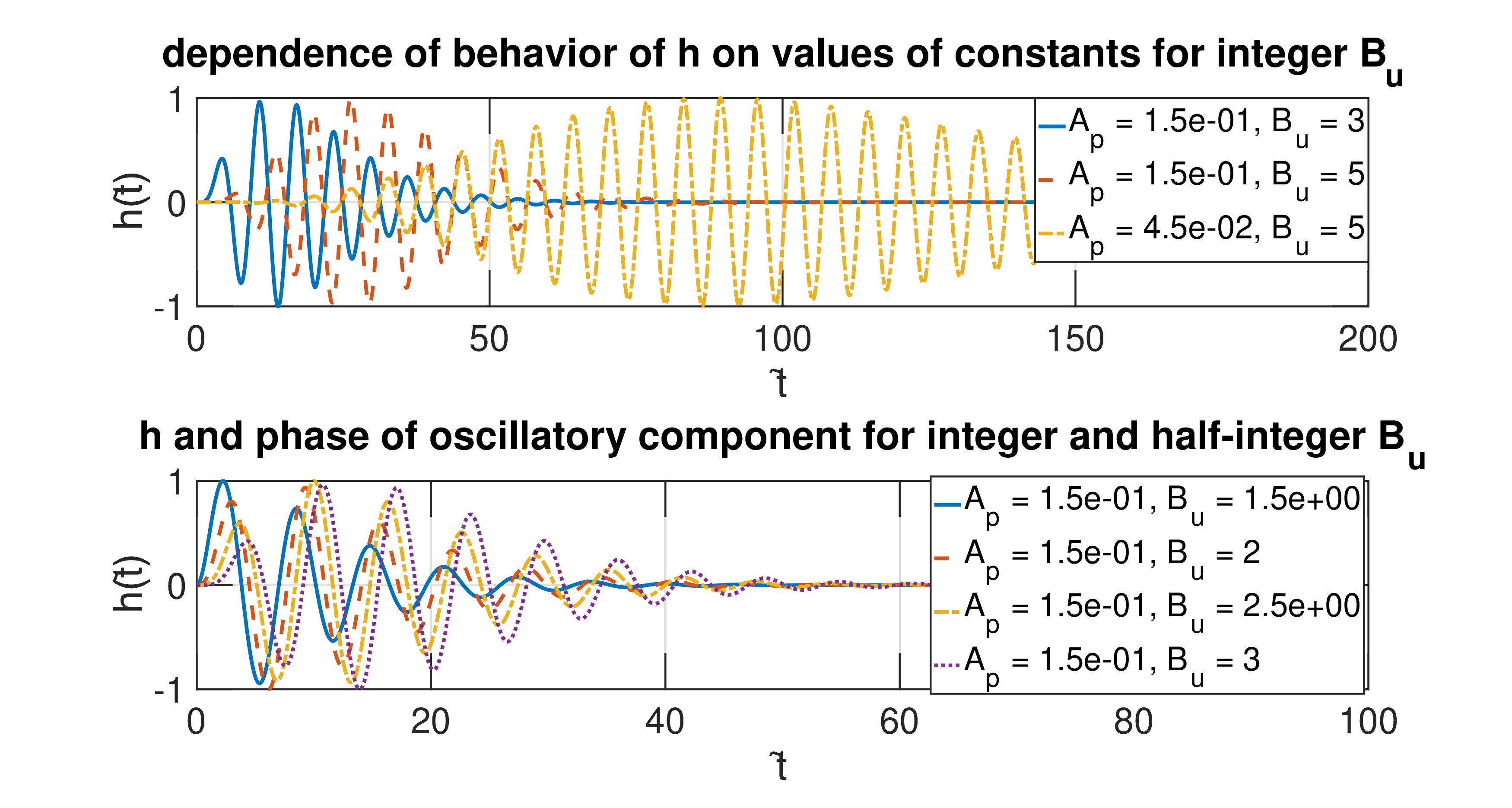}
    \caption{\textbf{Normalized impulse responses as a function of normalized time $\mathbf{\ttilde = 2\pi \text{CF}(x) t}$ with $\mathbf{\bp=1}$.} Impulse responses are plotted as normalized to the maximum absolute value. Top: shows the dependence of the impulse response (table \ref{tab:ImpulseResponseIntBuExpression}) behavior on the filter constants. For instance, delay increases with increasing $\Bu$ and decreasing $\Ap$. The figure is generated by varying $\Ap$ and \textit{integer} $\Bu$ (which limits the oscillatory component of $h(\ttilde)$ to a sin or cos). Bottom: shows that $h(\ttilde$) for integer and half-integer $\Bu$ (table \ref{tab:ImpulseResponseHalfIntBuExpression} or equation \ref{eq:hIntegerHalfinteger}) allows for greater control and variation of the phase of the oscillatory component rather than it being restricted to sin and cos as is the case for integer $\Bu$. }
    \label{fig:hDependenceOnParamsAndHintVsHalfIntPhase}
\end{figure}

Here we derive the $h(\ttilde)$ of the impulse responses, $g$, for the cases of integer $\Bu$. We do not discuss $\Bu = 1$ as it does not provide the desired envelope structure and instead corresponds to an impulse response with $h(\ttilde) = \frac{\expn{-\Ap \ttilde} \sin (\bp \ttilde)}{\bp}$, that will decay without a prior rise - which is inconsistent with specifications for signal processing applications and cochlear measurements. In table \ref{tab:ImpulseResponseIntBuExpression}, we present $h(\ttilde)$ for various integer values of $\Bu$, and plot two such instances in the top panel of figure \ref{fig:hDependenceOnParamsAndHintVsHalfIntPhase}.

\begin{table*}[htbp]
    \centering
    \begin{tabular}{|l|l|}
    \hline
    $\mathbf{\Bu}$ & $\mathbf{h(\ttilde) = }$ \\
    \hline
    $2$ & $\frac{1}{2\bp^3} \expn{-\Ap \ttilde}\bigg(\sin(\ttilde_b) - \ttilde_b \cos(\ttilde_b) \bigg)\xrightarrow[\text{approx.}]{\text{highest-order}} \frac{1}{2\bp^3} \expn{-\Ap \ttilde}\bigg( - \ttilde_b \cos(\ttilde_b) \bigg) $ \\
    \hline
    $3$ & $\frac{1}{8\bp^5} \expn{-\Ap \ttilde} \bigg(3\sin(\ttilde_b) - 3 \ttilde_b \cos(\ttilde_b) - \ttilde_b^2 \sin(\ttilde_b)) \bigg) \xrightarrow[\text{approx.}]{\text{highest-order}} \frac{1}{8\bp^5} \expn{-\Ap \ttilde} \bigg(- \ttilde_b^2 \sin(\ttilde_b)) \bigg) $ \\
    \hline
    $4$ & $\frac{1}{48\bp^7} \expn{-\Ap \ttilde} \bigg(15\sin(\ttilde_b) - 15 \ttilde_b \cos(\ttilde_b) - 6 \ttilde_b^2 \sin(\ttilde_b) + \ttilde_b^3 \cos(\ttilde_b) ) \bigg) \xrightarrow[\text{approx.}]{\text{highest-order}} \frac{1}{48\bp^7} \expn{-\Ap \ttilde} \bigg( \ttilde_b^3 \cos(\ttilde_b) ) \bigg)$ \\
    \hline
    $5$ & $\frac{1}{384\bp^9} \expn{-\Ap \ttilde} \bigg(105\sin(\ttilde_b) - 105 \ttilde_b \cos(\ttilde_b) - 45 \ttilde_b^2 \sin(\ttilde_b) + 10 \ttilde_b^3 \cos(\ttilde_b) + \ttilde_b^4 \sin(\ttilde_b) ) \bigg)  \xrightarrow[\text{approx.}]{\text{highest-order}} \frac{1}{384\bp^9} \expn{-\Ap \ttilde} \bigg( \ttilde_b^4 \sin(\ttilde_b) ) \bigg) $ \\
    \hline
    \end{tabular}
    \caption{$h(\ttilde)$ expressions, including the scaling coefficient, for several cases of positive integer values of $\Bu$ obtained from the inverse transform of the transfer function. These may also be generated from equation \ref{eq:hIntegerHalfinteger}.}
    \label{tab:ImpulseResponseIntBuExpression}
\end{table*}

Specifically, in the top panel of figure \ref{fig:hDependenceOnParamsAndHintVsHalfIntPhase}, we have shown the impulse response for two different values of the filter constant $\Bu$ which controls the phase of the oscillatory factor. The two illustrated cases have opposite polarity with respect to one another at the same $\ttilde_b$ \footnote{From a cochlear mechanics perspective, we note that in experimentally measured basilar membrane click responses, all the peaks measured from different locations along the cochlea approximately align in the same direction \cite{guinan}. Therefore, this matching polarity (after accounting for scaled time) may also be used to determine possible $\Bu$, as well its variation along the length of the cochlea, and consequently, filter constant values for filterbank applications. }. The polarity pattern for integer-exponent GEFs is limited to one of two choices - for odd and even $\Bu$ respectively, thereby limiting possible filter behavior.

\subsection{Integer-$\Bu$: Approximation as Realizable GTFs}

Table \ref{tab:ImpulseResponseIntBuExpression} shows that for any instance of the filter in which $\Bu \in \mathbb{Z} > 1$, $h(\ttilde)$ is the sum of gammatone filters each of which has an envelope that rises then decays and an oscillatory factor with tonal frequency $\bp \CFx$. The envelope of this response initially grows when the polynomial portion dominates, then decays when the exponential decay factor $\expn{-\Ap \ttilde}$ dominates the behavior.

\begin{figure}[htbp]
    \centering
    \includegraphics[width = \linewidth]{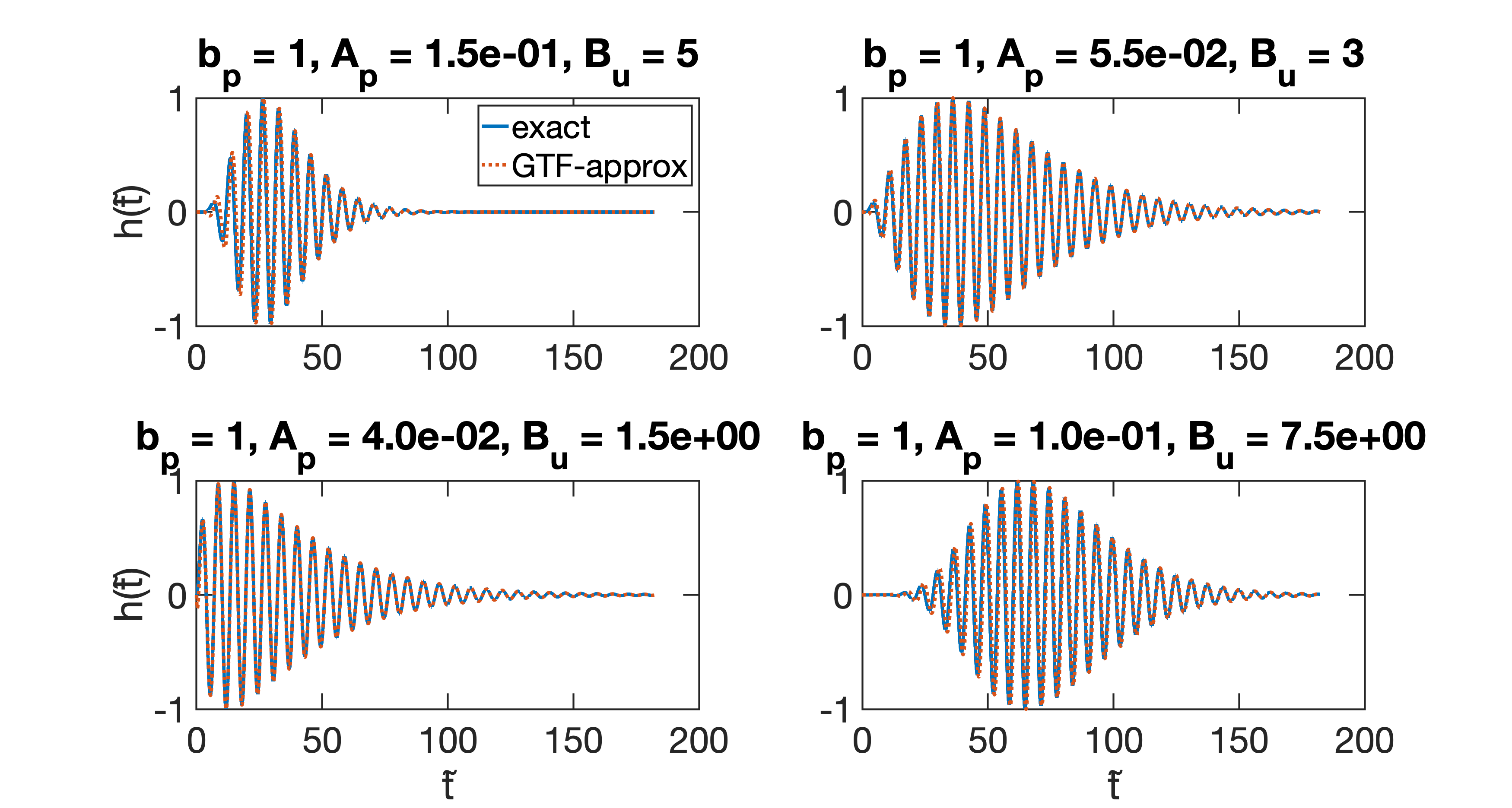}
    \caption{ \textbf{Comparison between the exact expression for impulse responses and its gammatone approximation:} For various values of filter constants, the figure shows the comparison between $h(\ttilde)$ (solid blue lines) as a function of normalized time $\ttilde = 2\pi \text{CF}(x) t$ with $\bp=1$ and its highest order term approximation, $\hGTF$ of equation \ref{eq:hGTFapprox} (dotted red lines), which is simply a single gammatone filter (GTF) extrapolated to allow for half-integer $\Bu$ and with a tonal component $\cos(\ttilde_b  - \Bu \frac{\pi}{2})$. The top figure panels is for integer $\Bu$, and the bottom panels are for half-integer-$\Bu$. As demonstrated in the figures, the impulse response is well-approximated by the highest-order term for small $\Ap$ except at the very earliest times.}
    \label{fig:hExactGammatoneComparison}
\end{figure}

\begin{figure}[htbp]
    \centering
    \includegraphics[width = \linewidth]{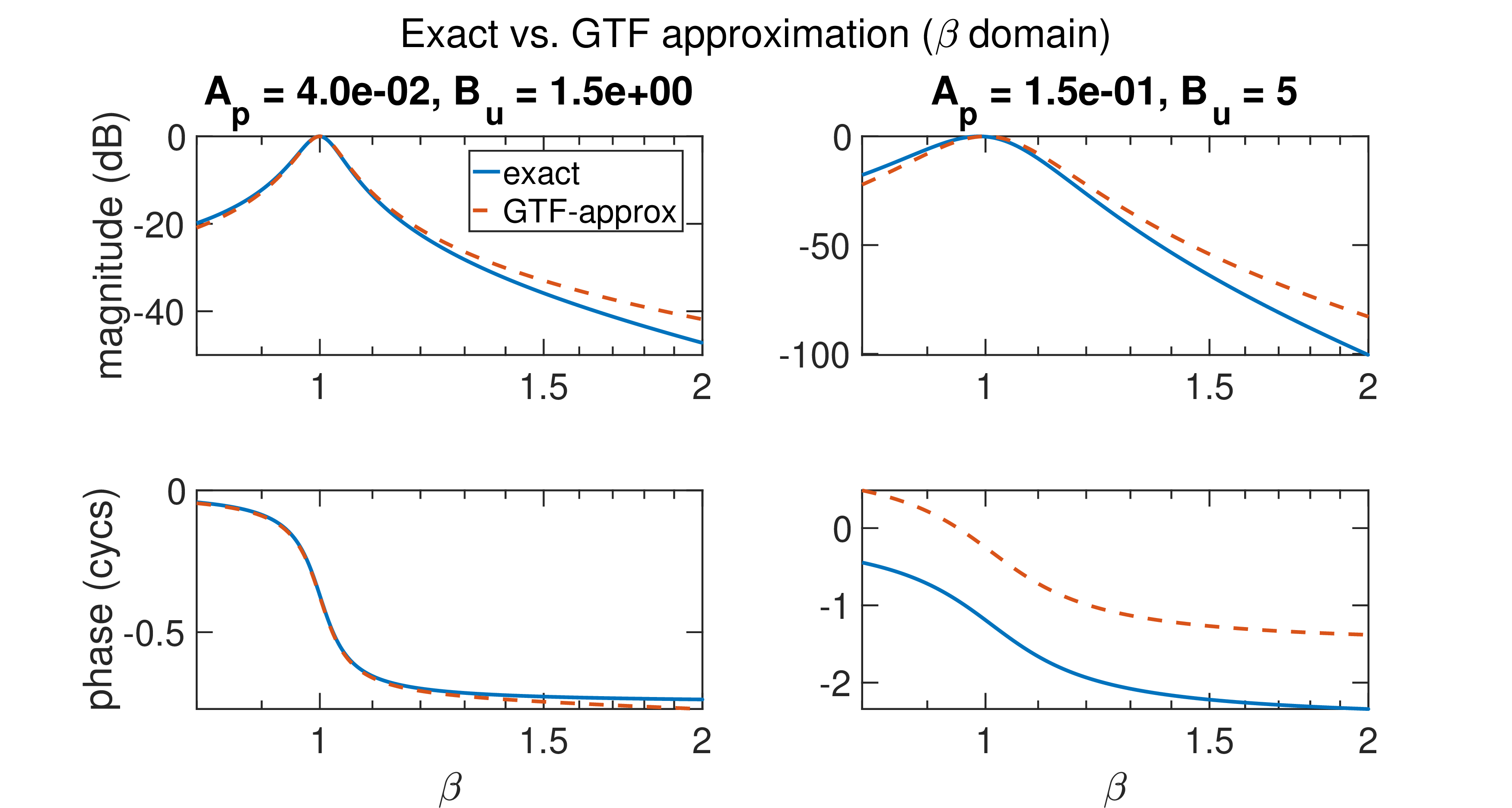}
    \caption{ \textbf{Comparison between the TFs for the exact expression for impulse responses and its approximation using extrapolated gammatone filters:} For a couple of values of filter constants used in figure \ref{fig:hExactGammatoneComparison}, the figure shows the comparison between the corresponding transfer functions for $h(\ttilde)$ (solid blue lines) and its highest order term approximation, $\hGTF$ (dotted red lines) which may be considered an extrapolation of the widely-used GTFs to non-integer $\Bu$. Comparing this figure with figure \ref{fig:hExactGammatoneComparison} shows that despite the exact and approximate impulse responses only differing at earlier times, their transfer functions \trackchanges{may differ outside the peak region for certain values of filter constants as particularly seen in the right panels of this figure.}}
    \label{fig:hExactGammatoneComparisonTFs}
\end{figure}

As time increases, the polynomials in parentheses become increasingly dominated by the highest order terms. Therefore, if $\Ap$ (which determines the decay rate) is sufficiently small, the expressions are dominated by the higher order terms except at the very initial times, and we may approximate the impulse responses using a single gammatone filter. Practically, this approximation is appropriate for $\Ap < 0.4$, which is always the case for the applications of interest. We illustrate that the exact and approximate impulse responses are similar beyond initial times for this set of filter constant values in the top panels of figure \ref{fig:hExactGammatoneComparison} - and illustrate that this similarity occurs despite the fact that the corresponding transfer functions (figure \ref{fig:hExactGammatoneComparisonTFs}) may show differences outside the peak region.

We may therefore approximate, $h(\ttilde)$ using only the exponential factor and the highest order polynomial term. Hence, for $\bp = 1$, we can write, 

\begin{equation}
    h(\ttilde) \appropto \expn{-\Ap \ttilde} \ttilde_b^{\Bu - 1} \cos(\ttilde_b - \Bu \frac{\pi}{2})\;, \quad \text{with } \Bu \in \mathbb{Z}>1 \;.
    \label{eq:GTFapprox}
\end{equation}

Consequently, $g(x,t)$ may be approximated as a Gamma-Tone-Filter (GTF) \footnote{which was developed purely from a filter perspective and has no bases in cochlear physics}. The GTF has a representation $a t^{n-1} \expn{-2\pi b t} \cos(2\pi f t + \phi)$  \cite{slaney1993efficient}, and is the basis on which a variety of filters in the gammatone family were developed, and has several architectures developed for different implementations.


\subsection{Parameterization}
From our expressions for impulse responses, and as illustrated in the top panels of figure \ref{fig:hDependenceOnParamsAndHintVsHalfIntPhase}, we may make the following comments regarding filter impulse response behavior and its dependence on filter constants:

\begin{itemize}
    \item $\bp$ specifies the tonal frequency of the impulse response. Specifically, this is $f = \frac{\ttilde_b}{2\pi t} = \text{CF}(x)\bp$ as can be seen from the arguments of the $\sin, \cos$. When $\bp = 1$, the tonal frequency is simply the peak frequency.
    \item $\Ap$ determines the decay of the envelope of the impulse response. This can be determined from the later parts of the response. Specifically, the decay factor at large times is, $ = \expn{-\Ap \ttilde} = \expn{-\Ap 2\pi \text{CF}(x) t}$. This means that for larger $\Ap$, the decay at larger times is faster than it is for smaller $\Ap$. In addition, the decay at larger times is faster for filters with higher peak frequency, \trackaddition{CF}. When expressed in terms of period of the CF, the decay time is constant assuming the same set of filter constant values. 
    \item $\Bu$ specifies the complexity of the impulse response. Specifically, it increases the number of polynomial terms, $\ttilde_b^k$ and the degree of the polynomial.
    \item $\Bu$ determines the phase shift of the oscillatory factor of the impulse response. This limits the polarity for integer-$\Bu$.
    \item Both $\Bu$ and $\Ap$ specify the time at which the envelope of the impulse response peaks which occurs at $\ttilde = \frac{\Bu - 1}{\Ap}$. For filters with the same set of filter constant values $\Ap, \Bu$, the time, $t$, at which the envelope peaks is inversely proportional to $\text{CF}(x)$. 
\end{itemize}

\subsection{Filter Design and Leveraging Efforts for Implementation}
As mentioned previously, the impulse responses are useful for deriving filter parameterizations and associated filter design criterion in terms of desired (or observed) time domain specifications for the impulse response and step response - e.g. settling time. The impulse response representation enables us to leverage several existing efforts developed for the widely used GTFs and other related filters towards developing analog and digital circuit implementations \cite{ngamkham2010analog}, \cite{onizawa2017area}.


\subsection{Generalization to Half-Integer-$\Bu$: Formulation}

Here we generalize our expressions for the GEF impulse response to half-integer-$\Bu$ cases in order to expand the variety of behavior which the filters may exhibit. Once we reduce the integer restriction on $\Bu$ to $\Bu = \frac{n}{2}$ with $n \in \mathbb{Z}^+$, we find that we may lessen the restriction on the values of $\Bu$ which result in desirable impulse response behavior. The  condition for having an $h(\ttilde)$ that has an envelope that  grows then decays becomes $\Bu \geq \frac{3}{2}$ rather than the more stringent condition required when $\Bu$ is restricted to integer values ($\Bu \geq 2$).

For integer and half-integer values of $\Bu$, we find that we may express the $h$ of the impulse response as,

\begin{equation}
    h(\ttilde) \propto \expn{-\Ap \ttilde} \ttilde^{\Bu - \frac{1}{2}} J_{\Bu - \frac{1}{2}}(\ttilde_b) \;,
    \label{eq:hIntegerHalfinteger}
\end{equation}

which is a generalization appropriate for both integer and half-integer-$\Bu$ \footnote{We expect that rational-$\Bu$ filters more generally can be expressed exactly by special functions that are a generalization of Bessel functions of the first kind.}. The above expression consists of: a Bessel function of the first kind (which is a decaying oscillatory function), a decaying exponential factor, and a growth factor due to polynomials of $\ttilde$. We illustrate this in figure \ref{fig:hDependenceOnParamsAndHintVsHalfIntPhase} (bottom panel), and provide the expressions including the scaling factors for a few instances of $\Bu$ in table \ref{tab:ImpulseResponseHalfIntBuExpression}. \trackaddition{The half-integer value $\Bu = \frac{1}{2}$ is not appropriate as it results in $\expn{-\Ap \ttilde} J_o(\ttilde_b)$ in which the envelope only decays contrary to the desired behavior of the filters for most applications}.

\begin{table}[htbp]
    \centering
    \begin{tabular}{|l|l|}
    \hline
    $\mathbf{\Bu}$ & $\mathbf{h(\ttilde) = }$ \\
    \hline
    $\frac{3}{2}=1.5$ & $\frac{1}{\bp} \expn{-\Ap \ttilde} \ttilde J_1(\ttilde_b)$ \\
    \hline
    $\frac{5}{2}=2.5$ & $\frac{1}{3\bp^2} \expn{-\Ap \ttilde} \ttilde^2 J_2(\ttilde_b)$ \\
    \hline
    $\frac{7}{2}=3.5$ & $\frac{1}{15 \bp^3} \expn{-\Ap \ttilde} \ttilde^3 J_3(\ttilde_b)$ \\
    \hline
    $\frac{9}{2}=4.5$ & $\frac{1}{105 \bp^4} \expn{-\Ap \ttilde} \ttilde^4 J_4(\ttilde_b)$ \\
    \hline
    \end{tabular}
    \caption{$h(\ttilde)$ expressions for several cases of positive half-integer values of $\Bu$. Note that the Bessel function of the first kind, $J_{\nu}(\ttilde_b)$ provides decaying oscillatory terms.}
    \label{tab:ImpulseResponseHalfIntBuExpression}
\end{table}

\subsection{Approximation as Extrapolated-GTFs}
As may be anticipated from the similarity in behavior between integer-exponent GEFs and GTFs, we find that we may \trackaddition{extend} the approximation using GTFs to cases with half-integer-$\Bu$ (as shown in figure \ref{fig:hExactGammatoneComparison}), and consequently, we extrapolate the term GTFs to include non-integer-$\Bu$,

\begin{equation}
    h(\ttilde)  \appropto \hGTF(\ttilde)
\end{equation}

\begin{equation}
    \hGTF \propto \expn{-\Ap \ttilde} \ttilde^{\Bu - \frac{1}{2}} \cos(\ttilde_b - \Bu \frac{\pi}{2})\;.
    \label{eq:hGTFapprox}
\end{equation}

We further expect that $h_{GTF}$ is a good approximation of GEFs with real $\Bu$ and not just integer and half-integer-$\Bu$. Once we remove the restriction of integer $\Bu$, we can achieve a tonal component that is not only restricted to $\sin(\ttilde_b)$ or $\cos(\ttilde_b)$, but rather may have a larger variety of phases. This is particularly clear from equation \ref{eq:hGTFapprox} \trackaddition{and illustrated in} figure \ref{fig:hDependenceOnParamsAndHintVsHalfIntPhase}.

The approximation of half-integer $\Bu$ GEFs as extrapolated GTFs is quite useful towards implementations. This is due to the fact that several implementations of the widely-used GTFs and related filters have been presented, and we may benefit from these towards developing realizable non-integer GEF implementations using the aforementioned approximations.

\section{ODEs for Integer $\Bu$}
\label{s:tODE}

\subsection{ODE Representation}

Here we derive the ODE representation for GEFs. Transforming equation \ref{eq:Pbold} for $\frac{P(x,\w)}{C(x)}$ into the time domain yields a linear ODE of order $2\Bu$,

\begin{equation}
   \bigg(\frac{d^2}{d\ttilde^2} + 2\Ap \frac{d}{d\ttilde} + |p|^2 \bigg)^{\Bu} q(x,t)  =  u(t) \;,
   \label{eq:tODE}
\end{equation}

where $v_{st}(\w) \xleftrightarrow{} u(t)$, $\frac{P(x,\w)}{C(x)} \xleftrightarrow{} q(x,t)$, and $\frac{d}{d\ttilde} = \frac{1}{2\pi\CFx} \frac{d}{dt}$. Neglecting matters of computational expense, the ODE may be formulated and simulated in a variety of ways and may be run in real time. 



\subsection{State-Space Formulation}
One approach for simulations (which we use to generate the responses in figure \ref{fig:equivBuInteger}) is to express the problem as a state-space formulation translated from the ODE of equation \ref{eq:tODE} (or the TF of equation \ref{eq:Pbold}) and then use Runge-Kutta based ODE solvers which are quite accurate. \trackchanges{For instance, we may express the state-space representation for a filter with a peak frequency $\textrm{CF}(x_i)$ as follows for an example case with $\Bu = 2$}. As we're interested in computing the output of a single filter rather than of a filterbank here, we may simply consider the variable in terms of $\ttilde_i$ rather than $t$ and $x$, and accordingly, we may simulate in $\ttilde_i$. Hence, we use, $\mathtt{u}(\ttilde_i) = u(\frac{\ttilde_i}{2\pi\mathrm{CF}(x_i)})$ and $\mathtt{q}(\ttilde_i) = q(t,x_i)$.

\begin{equation}
\begin{aligned}
\frac{d}{d\ttilde_i}\Vec{z}(\ttilde_i) & = 
\begin{bmatrix} 
&1&& \\
&&1& \\
&&&1 \\
-4\Ap&-4\Ap^2 - 2|p|^2&-4\Ap|p|^2& -|p|^4
\end{bmatrix} 
\Vec{z}(\ttilde_i) \\
& + 
\begin{bmatrix}
    0\\0\\0\\1
\end{bmatrix}\mathtt{u}(\ttilde_i) ,
 \quad \Vec{z}(0) = \Vec{0} \;,
\end{aligned}
\end{equation}

The first state variable in $\vec{z}$ is the filter output $\mathtt{q}(\ttilde_i)$. As we solve the state-space representation numerically, this representation is essentially in discrete time due to the nature of simulations and numerical ODE solvers.

\subsection{Properties}

The ODE representation is particularly easy and direct, and appropriate for software implementations for certain classification applications and perceptual studies. Furthermore, appropriate state-space formulations of the ODE representation may be used to determine equivalent orthonormal ladder filters for which circuit design procedure is easily automated \cite{johns1989orthonormal}.

Additionally, we may directly modify the ODE to incorporate additional behavior. One example is incorporating nonlinearity directly by making the coefficients dependent on $q$, rather than using the existing compression schemes. The ODE representation may also directly be modified to incorporate other behavior (such as memory, rectifier, refractory period, and various other mechanical and neural effects to better mimic auditory system processing) potentially by extensions to deterministic and stochastic delay-differential equations \cite{rameh2020single, goldwyn2011stochastic, rihan2018applications, botti2015delay}.

Despite its suitability for simple direct implementations and for incorporating additional behavior, the ODE representation is the most constraining in terms of possible values of $\Bu$ as our formulation restricts it to integer values. We may be able to extend the ODE representation to non-integer-$\Bu$ cases using fractional calculus methods but this is beyond the scope of this paper. While the ODE formulation presented here is itself restricted to integer $\Bu$, we use it in the next section, to derive our integral formulation which is not limited to the integer-$\Bu$ case.


\section{Integral Expressions}
\label{s:integralExpress}

Here, we take another approach for deriving a formulation for real-time processing - one that involves integration. We develop the integral representation using differential operator theory \cite{cheng2007advanced} which we extend to rational-exponent filters. In what follows, we first derive the integral representation for integer-$\Bu$ filters, then present the integral representations for the rational-$\Bu$ case.

\subsection{Integer $\Bu$}

Defining the differential operator,
\begin{equation}
    \Dtilde  \triangleq \frac{d}{d\ttilde} = \frac{1}{2\pi \text{CF}(x)}\frac{d}{dt} \;,
\end{equation}

we may express, $q(x,t)$ of equation \ref{eq:tODE} as,
\begin{equation}
    q(x,t; \Bu, \Ap, \bp; u(t))   = \bigg( \frac{1}{(\Dtilde-p)(\Dtilde-\pconj)} \bigg)^{\Bu} u(t)  \;.
    \label{eq:tInvODE}
\end{equation}

As we are interested in the solution with zero initial conditions in response to any smooth function $u(t)$ \footnote{We note that equation \ref{eq:myDOT} does not correspond to the zero-initial-condition solution for inputs that are impulses and other generalized functions.}, let us define our integration operator as,

\begin{equation}
    \frac{1}{\Dtilde_o} f(\ttilde) \triangleq \defint{0}{\ttilde}{f(\ttilde')}{\ttilde'}  = 2\pi \text{CF}(x) \defint{0}{t}{g(t')}{t'} \;.
    \label{eq:myDOT}
\end{equation}

While we have deviated from the classical formulation of differential operator theory, dealing with these definite integrals provides the analytic expressions for the desired solutions directly rather than dealing with an indefinite integral then solving for the constants using initial conditions. 


Using differential operator theory and the above formula for integration, we may express the response of equation \ref{eq:tInvODE} as follows:

\begin{equation}
\begin{aligned}
     q(x,t; \Bu, \Ap, \bp; u(t))  & = \bigg( \frac{1}{(\Dtilde_o-p)(\Dtilde_o-\pconj)} \bigg)^{\Bu}  u(t)\\
    & = \bigg( \frac{1}{\Dtilde_o-\pconj} \bigg)^{\Bu} \bigg( \frac{1}{\Dtilde_o-p} \bigg)^{\Bu}  u(t) \\ 
    & = \expn{\pconj \ttilde} \bigg( \frac{1}{\Dtilde_o} \bigg)^{\Bu} \expn{-\pconj \ttilde} \expn{p \ttilde} \bigg( \frac{1}{\Dtilde_o} \bigg)^{\Bu} \expn{-p \ttilde} u(t) \\
    & = \expn{\pconj \ttilde} \bigg( \frac{1}{\Dtilde_o} \bigg)^{\Bu} \expn{2i\bp \ttilde} \bigg( \frac{1}{\Dtilde_o} \bigg)^{\Bu} \expn{-p \ttilde} u(t) \;,
\end{aligned}
\label{eq:tint}
\end{equation}

which, despite its intermediate operations, results in a real output for any given real input. As an example, for the case of $\Bu = 1$, this yields,

\begin{equation}
\begin{aligned}
  q(x,t; 1, \Ap, \bp; u(t))  =  \bigg( 2\pi\CFx \bigg)^ 2 \expn{2\pi\CFx \pconj t} \times \\
  \defint{0}{t}{\expn{4\pi\CFx i\bp t'}  
  \defint{0}{t'}{\expn{-2\pi \CFx p t''} u(t'')}{t''}}{t'} \;,
\end{aligned}
\end{equation}

For more reasonable values of integer $\Bu$, each $\big( \frac{1}{\Dtilde_o} \big)^{\Bu}$ of equation \ref{eq:tint} results in nested integrals,

\begin{equation}
    \bigg( \frac{1}{\Dtilde_o} \bigg)^{\Bu} f(\ttilde) =  \defintrev{0}{\ttilde}
    {
        \defintrev{0}{\ttilde'}
        {
            \dots \defintrev{0}{\ttilde^{(\Bu-1)}} {f(\ttilde^{(\Bu)})}
            {\ttilde^{(\Bu)} }
        }{\ttilde''}
    }{\ttilde'}  \;.
    \label{eq:nested}
\end{equation}

It is possible to compute solutions $q$ for a given input $u$, for the case of integer $\Bu$, directly using equations \ref{eq:tint} and \ref{eq:nested}. However, it is cumbersome, and has several nested integrals - the number of which is based on the value of $\Bu$ which can be quite large. 

We reduce these operations into only two integrals using the Cauchy formula for repeated integration which is valid for integer values of $\Bu$ \cite{lovoie1976fractional, loverro2004fractional}. This allows us to express equation \ref{eq:tint} using,
\begin{equation}
     \bigg( \frac{1}{\Dtilde_o} \bigg)^{\Bu} f(\ttilde) = \frac{1}{(\Bu - 1)!} \defint{0}{\ttilde}{(\ttilde - \tau)^{\Bu - 1} f(\tau)}{\tau}\;,
\end{equation}

which allows us to derive the following equivalent to equation \ref{eq:tint}  that reduces the expression for the output to only two nested integrals \textit{regardless} of the value of $\Bu$, 

\begin{equation}
\begin{aligned}
    \mathtt{q}(\ttilde_i; \Bu, \Ap, \bp; \mathtt{u}(\ttilde_i)) =  \bigg(\frac{1}{(\Bu - 1)!}\bigg)^2 \expn{\pconj \ttilde_i}  \times 
    \\
    \int_{0}^{\ttilde_i} {d\tau}
        (\ttilde_i - \tau)^{\Bu - 1} \expn{2i\bp \tau}
            \int_{0}^{\tau}{dT}{
            (\tau - T)^{\Bu - 1} \expn{-p T} \mathtt{u}(T)
            }
            \;.
    \label{eq:integralInteger0t}
\end{aligned}
\end{equation}


While it is appropriate to express this in the $t$ domain, for a given filter we express this in the $\ttilde_i$ domain for compactness. We use the notation $\mathtt{q}$ for $\mathtt{q}(\ttilde_i) = q(x_i,t)$ and $\mathtt{u}$ for $\mathtt{u}(\ttilde_i) = u(t(\ttilde_i, x_i)) = u(\frac{\ttilde_i}{2\pi\textrm{CF}(x_i)})$.


For any input that is a smooth function \footnote{If $u(t) = \delta(t)$, the expressions result in the full solution with all initial conditions being zero with the exception of the initial condition corresponding to the highest derivative which is unity. Notice that using the integral representation with $u(t) = \delta(t)$ directly results in our expressions for the impulse responses $g(x,t) = 2\pi\CFx h(\ttilde)$. This can most easily be checked with $\Bu = 1$.}, equation \ref{eq:integralInteger0t} may be used to compute the full solution (including both the complementary and particular solutions) when the initial conditions are all zero.

In the above expression, a parameter of the integrand also appears as an upper limit of integration. This is no issue for numerical computation of the output. However, for the analytic study of the output of certain signals, we also provide an alternative equivalent expression by using,

\begin{equation}
     \defint{0}{y}{f(z;y)}{z} = y \defint{0}{1}{f(Zy;y)}{Z} \;,
\end{equation}

where we perform the change of variables from $z$ to $Z$ such that $z = Zy$, resulting in a definite integral with a numerical value for the limits of integration. This results in the following expression for $\mathtt{q}(\ttilde_i)$ - where we have dropped the subscript $i$ to reduce clutter,

\begin{equation}
\begin{aligned}
    \mathrm{q}(\ttilde; \Bu, \Ap, \bp; \mathrm{u}(\ttilde)) =
    \\
    \bigg(\frac{1}{(\Bu - 1)!}\bigg)^2 \expn{\pconj \ttilde} 
    \int_{0}^{1}{d\zeta}
        \ttilde^{\Bu} (1 - \zeta)^{\Bu - 1} 
    \times \\
    \expn{2i\bp \ttilde \zeta}
            \int_{0}^{1}{d\chi}{ (\ttilde \zeta)^{\Bu}
            (1-\chi)^{\Bu - 1} \expn{-p \ttilde \zeta  \chi} \mathrm{u}(\ttilde \zeta   \chi)
            }
     \\
     = 
    \frac{\ttilde^{2\Bu}}{\bigg((\Bu - 1)!\bigg)^2 }\expn{\pconj \ttilde} 
    \int_{0}^{1}{d\zeta}
        (1 - \zeta)^{\Bu - 1} \zeta^{\Bu} 
        \times \\
        \expn{2i\bp \ttilde \zeta}
            \int_{0}^{1}{d\chi}{ 
            (1-\chi)^{\Bu - 1} \expn{-p \ttilde \zeta  \chi} \mathrm{u}( \ttilde \zeta  \chi)
            } \;.
    \end{aligned}
    \label{eq:integralInteger01}
\end{equation}


Both equations \ref{eq:integralInteger0t} and \ref{eq:integralInteger01} are integration-based expressions that may be used to evaluate the responses to a given input signal. Importantly, $\Bu$ occurs only as any other constant parameterizing the expression. Consequently, the same expression may be used regardless of the value of $\Bu$.

The integral representation is easily formulated for  software (including embedded software) realizations. Any integrator-based digital hardware realizations of the representation in this section must have the capacity of handling complex numbers as part of the intermediate operations.

\subsection{Rational $\Bu$}

Here, we extend the integral formulation of the filter to the case of rational exponents. In the previous section, we used the Cauchy formula for repeated integrals to derive a simple integral representation \trackaddition{for integer-$\Bu$ GEFs}. Here, we use its generalization to non-integers - the Riemann–Liouville integral in order to generalize our expression for $q$ to rational $\Bu$ for the case of zero initial conditions. Utilizing the Riemann–Liouville fractional integral operator \cite{lovoie1976fractional, loverro2004fractional} results in the following solution parallel to equations \ref{eq:integralInteger0t} and \ref{eq:integralInteger01}.

\begin{equation}
\begin{aligned}
    \mathrm{q}(\ttilde; \Bu, \Ap, \bp; \mathrm{u}(\ttilde)) = 
    \\ \bigg(\frac{1}{\Gamma(\Bu)}\bigg)^2 \expn{\pconj \ttilde} \int_{0}^{\ttilde}{d\tau}
        (\ttilde - \tau)^{\Bu - 1} 
        \times \\
        \expn{2i\bp \tau}
            \defintrev{0}{\tau}{
            (\tau - T)^{\Bu - 1} \expn{-p T} \mathrm{u}(T)
            }{T}
     \\
    = 
    \frac{\ttilde^{2\Bu}}{\bigg(\Gamma(\Bu)\bigg)^2 }\expn{\pconj \ttilde} \int_{0}^{1} d\zeta
        (1 - \zeta)^{\Bu - 1} \zeta^{\Bu} 
    \times \\
        \expn{2i\bp \ttilde \zeta}
            \int_{0}^{1} d\chi 
            (1-\chi)^{\Bu - 1} \expn{-p \ttilde \zeta  \chi} \mathrm{u}( \ttilde \zeta  \chi)
             \;.
\end{aligned}
\end{equation}


Note that since we deal with zero initial conditions, we need not concern ourselves with the various definitions for integral expressions for fractional derivatives.

\begin{figure}[htbp]
    \centering
    \includegraphics[width = \linewidth]{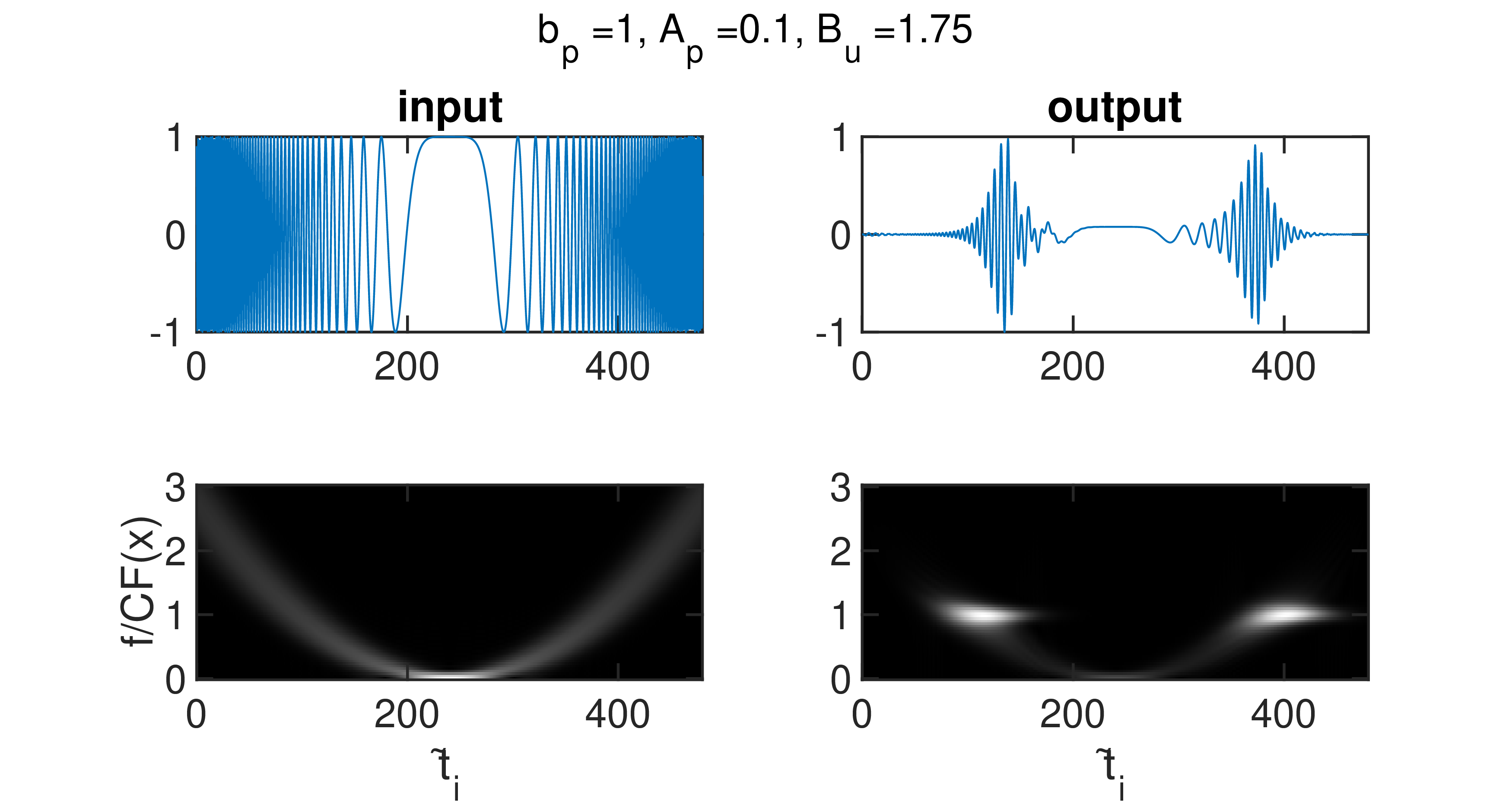}
    \caption{\textbf{The integral expressions can be used to compute responses for the case of rational-$\Bu$ GEFs: } The figure shows the normalized output, $q$ (top right) \trackchanges{of} any filter with peak frequency $\textrm{CF}(x_i)$. The output is in response to a quadratic chirp. The input $\mathtt{u}(\ttilde_i) \triangleq u(t(x_i,\ttilde_i))$, and the output $\mathtt{q}(\ttilde_i) \triangleq q(t,x_i)$ are plotted as a function of scaled time, \trackaddition{$\ttilde_i = 2\pi \textrm{CF}(x_i) t$}. The response is computed using the integral expression. The figure also shows the spectrogram of both the input (bottom left) and normalized output (bottom right). The plots show that filter primarily responds (with a time delay - dependent on $\Ap$ and $\Bu$ values) to instantaneous frequencies that are at, and near, the CF. It is relatively insensitive to frequencies further away from the CF. }
    \label{fig:integralForChirp}
\end{figure}

In figure \ref{fig:integralForChirp}, we demonstrate that the integral formulation is realizable by using it to compute the response to a quadratic chirp for the case of non-integer $\Bu$. The solution behaves as expected in that it responds maximally to frequencies close to the filter's peak frequency (with some delay).

\subsection{Properties}

We are unaware of any previous efforts towards integral representations for GEFs or any other filters. The integral representation can directly be extended to any all-pole filter. It is particularly useful to do so for classical filters and rational-exponent extensions that can be used in applications that may benefit from the real-time processing nature and other benefits afforded by the integral formulation.

The integral formulation allows for real-time signal processing and does not require buffering or preprocessing. It is in itself a  continuous time rather than a discrete time formulation, though may be discretized for various realizations. It provides expressions to be evaluated (rather than equations to be solved) for direct solutions to a given input $u(t)$. Consequently, it is computationally efficient if using efficient algorithms for integration. Additionally, as is the case with the extrapolated GTF-approximation for the impulse response representation, we may readily use existing methods developed for implementing the integer-exponent filters for the rational-exponent ones.

The outputs are evaluated via simple operations such as exponentiation and integration (or summation) which make it directly implementable in software, and allows for potential integrated digital circuit realizations that build on design and implementation in hardware description languages. The error for digital software implementations is due to the chosen integration scheme. There are no derivatives involved as the GEF is an all-pole filter. This is desirable as taking derivatives of noisy data should be avoided \footnote{We note that while it is not an all-pole filter, the $V$ filter of \cite{alkhairy2019analytic} can also be expressed without taking any derivatives as it only has a single zero corresponding to a first derivative of $u(t)$ which can be avoided by using integration by parts in the innermost integral.}.





\section{Equivalence of Representations}
\label{s:equivRep}


In this section, we test our derivations and demonstrate that the various representations are consistent with one another. We do this by using the various representations to compute the response of a filter to an input $\mathtt{u}(\ttilde)$ - where we have dropped the previously-used filter-specific subscript $i$ in $\ttilde_i$ for simplicity. We carry out the equivalency tests for: (1) a case of integer $\Bu$ to compare all representations, and (2) a case of half-integer $\Bu$ to compare all representations except for the ODE formulation which is limited to integer-$\Bu$. In both cases, we choose our input \trackaddition{signals} such that \trackaddition{both} the input and output \trackaddition{signals} have analytic expressions in the time and frequency domains. This allows us to generate exact analytic solutions which we then use \trackaddition{to test the accuracy of our formulations by comparing} the outputs computed using each of the formulations against \trackaddition{the true solution generated analytically}.

\subsection{Integer-Exponent}

\begin{figure}[htbp]
    \centering
    \includegraphics[width = \linewidth]{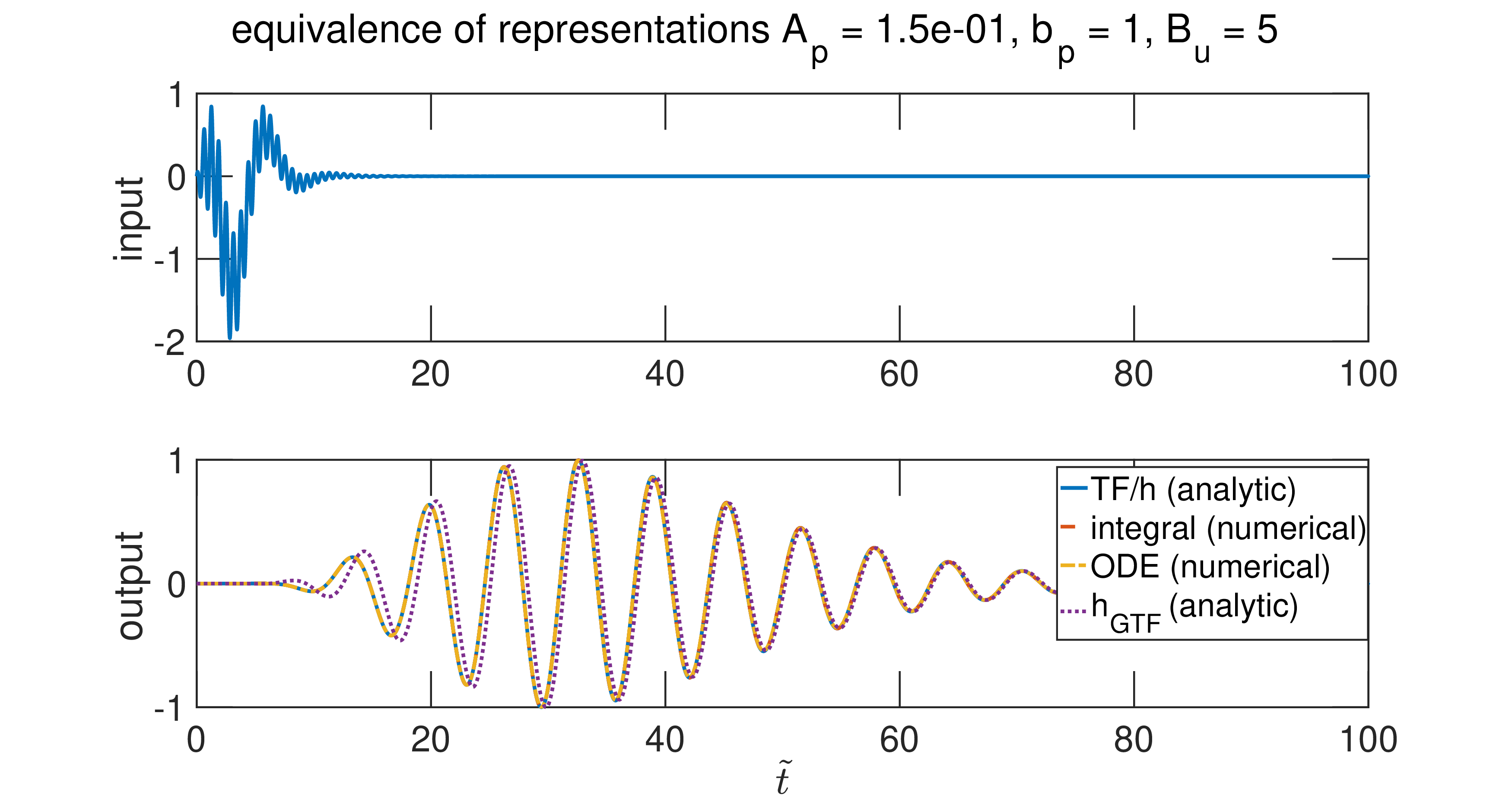}
    \caption{\textbf{The GEF representations are equivalent for the case of integer $\Bu$}: The figure shows the normalized output of a filter with a specified peak frequency $\textrm{CF}(x_i)$. \trackaddition{The input (top figure) is generated using equation \ref{eq:inputForIntegerBuComparison}}. The output, $\mathtt{q}(\ttilde_i)$ is plotted as a function of scaled time,  $\ttilde_i = 2\pi \textrm{CF}(x_i) t$, in response to the input. As we are interested in displaying the output of a single filter (as opposed to the outputs of several filters in a filterbank wherein each filter has its own peak frequency), we have represented the input and response in scaled time:  $\mathtt{u}(\ttilde_i) \triangleq u(t(x_i,\ttilde_i))$, and $\mathtt{q}(\ttilde_i) \triangleq q(t,x_i)$. The response (bottom figure) is computed numerically using 
    the integral representation (dashed red line) and the ODE representation with zero initial conditions (dash-dotted yellow line), and analytically using the TF or impulse response representation (solid blue line) as well as the GTF approximation (dotted purple line). The responses from the various representations are equivalent for integer $\Bu$, with the exception of the response generated using the GTF approximation which deviates from the exact response for the selected set of filter constant values (more so at earlier times).}
    \label{fig:equivBuInteger}
\end{figure}

For the integer-$\Bu$ case, we compare the outputs of the various methods in response to an input,

\begin{equation}
    \mathtt{u}(\ttilde) = \ttilde \cos(10\ttilde) \expn{-\frac{\ttilde}{2}} + \ttilde^3 \expn{-\ttilde} \cos(\ttilde)\;,
    \label{eq:inputForIntegerBuComparison}
\end{equation}

for which we may derive a true analytic solution. \trackaddition{In figure \ref{fig:equivBuInteger}, we show} the outputs generated using the various representations and compare \trackaddition{them} against the true analytic solution.

We also include the output generated using the $h_{GTF}$ approximation (generated analytically). The figure shows that, at least for integer-$\Bu$, all \trackchanges{\textit{exact}} GEF representations are equivalent. Additionally, the response generated using $h_{GTF}$ is qualitatively consistent with the exact solution. Quantitatively - for the set of filter constant values used in the figure, it shows only small deviations and only at earlier times. Consequently, differentiating between GEFs and GTFs (and other filters in the gammatone family) and restricting one's self to a particular filter is not particularly fruitful. Instead, we encourage processing using any of the filters in a manner best suited to selected implementations while extrapolating design and analysis methods of any related filters best suited to those tasks \cite{paperB1}.

\subsection{Half-Integer-Exponent}

\begin{figure}[htbp]
    \centering
    \includegraphics[width = \linewidth]{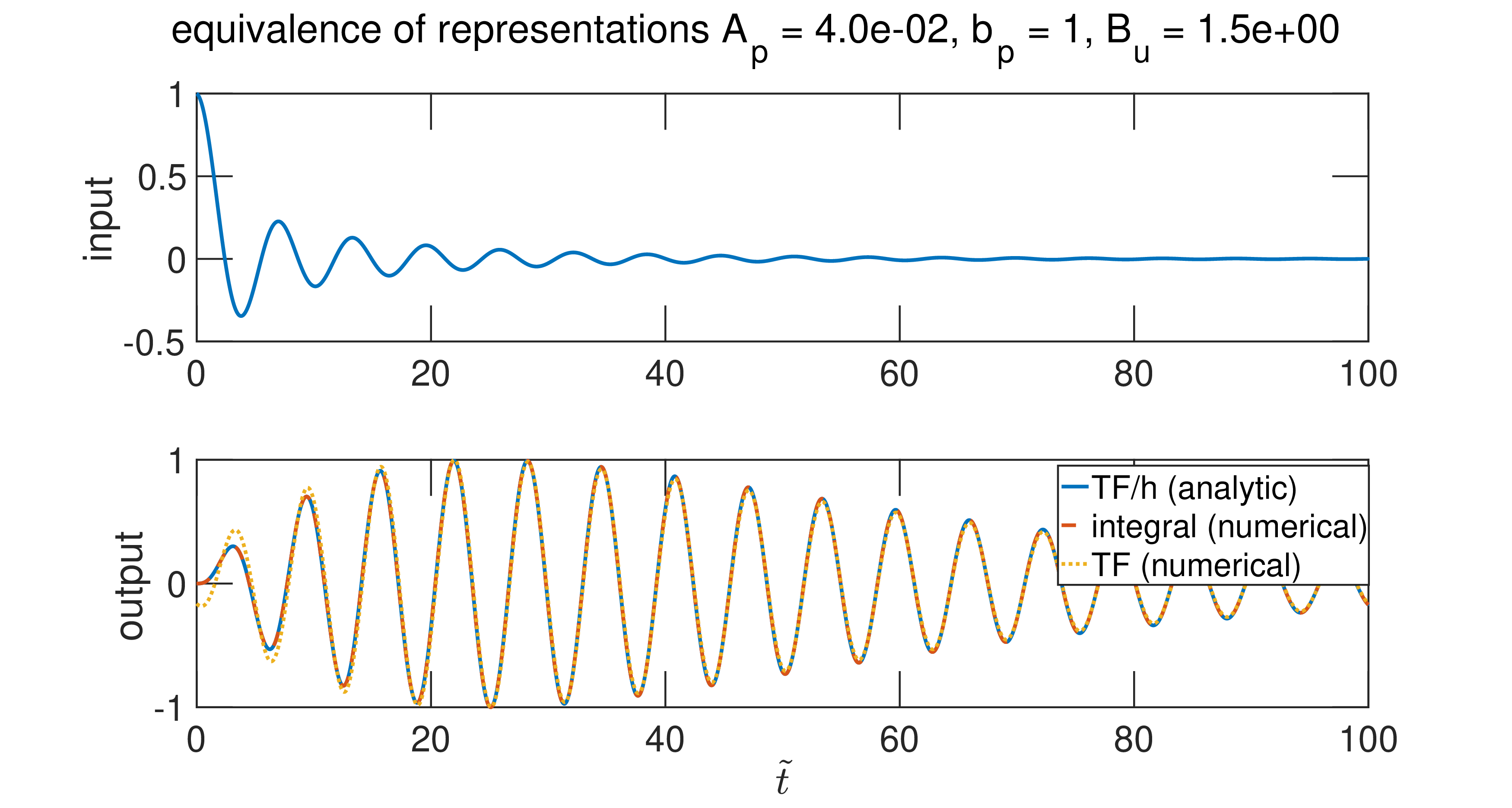}
    \caption{\textbf{The GEF integral formulation is the most accurate method used to compute the response for half-integer-exponent filters:} Similar to figure \ref{fig:equivBuInteger}, we plot the (normalized) output of a filter to an input (top figure) using the various representations for computing the output. However, in this figure, the filter has a half-integer-$\Bu$. The input is generated based on equation \ref{eq:inputForHalfIntegerBuComparison}. The output that is computed numerically using the integral representation (dashed red line) is equivalent to the analytic solution generated using the TF/h representations (solid blue line). On the other hand, the computational implementation of the TF representation (using FFTs) result in deviations from the true solution - which are particularly visible at earlier times. }
    \label{fig:equivRepsHalfInteger}
\end{figure}

In figure \ref{fig:equivRepsHalfInteger}, we compare the responses for the case of half-integer $\Bu$ to an input,

\begin{equation}
    \mathtt{u}(\ttilde) = \expn{-\Ap \ttilde} \ttilde^{a-\frac{1}{2}} J_{a - \frac{1}{2}}(\ttilde_b)  
    \;,
    \label{eq:inputForHalfIntegerBuComparison}
\end{equation}

\trackaddition{with $a = \frac{1}{2}$,} which has a transfer function that - neglecting the gain constant, is represented by $\big(s^2 + 2\Ap s + \Ap^2 + \bp^2 \big)^{-a}$. We choose the input to have the same base TF as the GEF (i.e. same $\Ap, \bp$ but any positive exponent) which allows us to analytically derive the true solution. We then compare against this solution to determine the accuracy of outputs computed using the various representations.

The outputs are generated analytically using the TF or impulse response representation and numerically using the integral representation. The figure shows that the output computed using the integral formulation is equivalent to that derived analytically. Additionally, we numerically compute the responses using the TF representation with FFTs/IFFTs and show that - in contrast to the outputs computed using the integral formulation, the output diverges from that derived analytically using the same representation. This may be a result of: additional numerical processing requirements for computing the response using TFs and FFTs/IFFTs; and the fundamental difference between the assumed nature of signals - in terms of being periodic and discrete, for which Discrete Fourier Transforms (used to generated the solution with FFT/IFFT algorithms) are appropriate compared to the assumed nature of the signals for which the Fourier Transforms (used to derive the true/analytic solution) are appropriate. This suggests that the integral expressions may be considered more desirable for software implementations in the sense that it does not require additional processing beyond evaluating the integral expression - and hence is more accurate.  

Consequently, for digital software implementations, we encourage the use of the integral formulation which is: accurate, useful for real-time realizations (if using efficient integration algorithms), appropriate for non-integer exponent GEFs, and does not require any preprocessing choices. These observations not only apply for software implementations of integer and rational-exponent GEFs, but also directly translate to other filters for which the base filter may be represented as all-pole rational transfer functions.

\section{Conclusions and Future Directions}
\label{s:conclusionFutureDir}

\subsection{Conclusions}

In this paper, we presented realizable rational-exponent GEFs. The rational-exponent filters are useful for an array of signal processing applications. They enable a continuum of behavior not accessible by the integer-exponent filters previously introduced, while being realizable and also simple to analyze in terms of causality and stability. Fractional-order filters, in contrast, are also designed to access  flexible filter behavior, but they are limited in terms of maximum filter order (and hence range of behavior) and it is more difficult to: analyze their causality and stability, to characterize behavior, and to formulate filter design methods.

We derived and discussed equivalent representations in the frequency and time domains - transfer functions, impulse responses, ODEs, and integral expressions. Each of the various representations is appropriate for different uses: e.g. characterization of additional flexibility of behavior accessible in the rational-exponent case; analysis of causality and stability; extensions to incorporate nonlinearity; filter-characteristics-based parameterization for filter design based on desired peak frequencies, quality factors, settling times, etc; identifying approximations in terms of extrapolations of widely-used filters which allows us to leverage existing architectures towards  various realizations; real-time processing requirements; and pre-processing specifications.

We have shown that outputs may be generated computationally using each of the formulations using software, and briefly addressed future extensions to hardware realizations. To test that we have properly derived the various representations and to assess their accuracy, we examined the responses to the same inputs using the various representations, and studied (in section \ref{s:equivRep})  the accuracy of the various representations used to generate outputs. \trackaddition{Open source code for rational-exponent GEFs may be found at \url{https://github.com/AnalyticModeling/GEFs}.}


\subsection{Future Directions}

\subsubsection{Extensions to Other Filters}

We have presented rational-exponent filters that consist of a second-order filter - as a base, raised to a rational number to access a wide range and continuum of filter behavior not classically achievable while also allowing for simple stability analysis and direct realizations. The fundamental methods that we used to derive the rational-exponent representations for this filter may be extended to derive other rational-exponent filters building on various types of base filters. A related example from existing work may be expressed in the form $(\frac{1}{s+\tau})^{\Bu}$ with a low-pass RC circuit $\frac{1}{s+\tau}$ for its base filter \cite{helie2014simulation}.  As the stability of any rational-exponent filter is based entirely on that of the base filter, we ideally recommend restricting the base filters to fourth-order filters (or otherwise higher order filters with special forms) such that pole placement to achieve a stable system is particularly simple and does not require numerical solutions.

\subsubsection{Applications and Implementations}

Future directions also involve pursuing some of the applications mentioned in section \ref{s:intro} using our filters that are relieved of the integer-$\Bu$ restriction. These applications generally make use of the GEFs in a parallel filterbank configuration rather than as individual filters, with each filter in the filterbank having a specified set of filter characteristics including peak frequency, quality factor, and group delay.  Some of these applications may benefit from deriving alternate parameterizations (e.g. based on frequency-domain characteristics), selecting filter constant values, developing architectures appropriate for certain implementations, or incorporating certain features such as nonlinearity.

The representations presented may be used quite directly to develop software realizations as eluded to in this paper. Future \trackchanges{directions} may involve work towards digital hardware realizations based on certain representations. Such efforts include those towards: (1) equivalent filter representations for the state-space formulation of the ODE representation that allow for easily automated circuit design; (2) integrator-based implementations of the integral representation; and finally - and most easily, (3) extrapolating several existing architectures developed for classical GTFs and related filters which serve as suitable approximations for rational-exponent GEFs upon extrapolation.

\subsubsection{Characteristics-Based Filter Design}

The transfer function representation allows for deriving parameterizations of GEFs and for formulating filter design paradigms based on desired sets of filter characteristics such as peak frequency, $Q_{3\textrm{dB}}$, and $Q_{erb}$ \cite{paperB1}. Similarly, the impulse response representation, and its integral, may be used for deriving parameterizations based on impulse response or step response characteristics such as settling time, instantaneous frequency, and functions of the envelope shape. Such parameterizations are useful for easily designing the filters based on desired behavior rather than using a fixed set of generic values of constants or optimization given the desired frequency response. Additionally, a combination of parametization using time and frequency domain characteristics provides a mapping between characteristics in the time domain and those in the frequency domain which would be useful for building intuition towards filter design and analysis.




\end{document}